\documentclass[prb,twocolumn,groupedaddress,superscriptaddress,nobibnotes]{revtex4-1}
\usepackage{eurosym}
\usepackage{amsfonts}
\usepackage{amsmath}
\usepackage{amssymb}
\usepackage{graphicx}
\usepackage{float}

\begin{document}

\title{Stiffnessometer, a magnetic-field-free superconducting stiffness meter and its application}

\author{Itay Mangel}
\email[E-mail: ]{itaymangel89@campus.technion.ac.il}
\affiliation{Department of Physics, Technion-Israel Institute of Technology, Haifa, 3200003, Israel}

\author{Itzik Kapon}
\affiliation{Department of Physics, Technion-Israel Institute of Technology, Haifa, 3200003, Israel}

\author{Nitzan Blau}
\affiliation{Department of Physics, Technion-Israel Institute of Technology, Haifa, 3200003, Israel}

\author{Katrine Golubkov}
\affiliation{Department of Physics, Technion-Israel Institute of Technology, Haifa, 3200003, Israel}

\author{Nir Gavish}
\affiliation{Department of Mathematics, Technion-Israel Institute of Technology, Haifa, 3200003, Israel}

\author{Amit Keren}
\email[E-mail: ]{keren@physics.technion.ac.il}
\affiliation{Department of Physics, Technion-Israel Institute of Technology, Haifa, 3200003, Israel}

\date{\today }

\begin{abstract}
	
We provide a detailed account for a new method to measure superconducting stiffness $\rho_{s}$, critical current density $j_c$, and coherence length $\xi$, in one apparatus, without subjecting the sample to magnetic field or attaching leads. The method is based on the London equation $\mathbf{j}=-\rho_{s}\mathbf{A}$, where ${\bf j}$ is the current density and ${\bf A}$ is the vector potential. Using a rotor free $\bf{A}$ and a measurement of $\bf{j}$ via the magnetic moment of a superconducting ring, we determine $\rho_{s}$. By increasing $\mathbf{A}$ until the London equation fails we determine $j_c$ and $\xi$. The method is sensitive to very small stiffness, which translates to penetration depth $\lambda \lesssim 1$~mm. It is also sensitive to low critical current density $j_c \sim 10^3$ Amm$^{-2}$ or long coherence length $\xi \sim 1$~$\mu$m. Naturally, the method does not suffer from demagnetization factor complications, the presence of vortices, or out-of-equilibrium conditions. Therefore, the absolute values of the different parameters can be determined. We demonstrate the application of this method to La$_{2-x}$Sr$_{x}$CuO$_{4}$ with $x=0.17$.

\end{abstract}

\maketitle

\section{Introduction} \label{Introduction}

Superconducting stiffness $\rho_s$ is defined via the gauge invariant relation between the current density $\bf{j}$, the vector potential $\bf{A}$, and the complex order parameter $\Psi  =  \psi({\bf r})  {e^{i\phi ({\bf{r}})}}$ , with $\psi({\bf r})\ge 0$, according to
\begin{equation}
{\bf{j}} = \rho_s \left( {\frac{\Phi_0}{2\pi} \nabla \phi  - {\bf{A}}} \right)
\label{JtoAandPhi}
\end{equation}
where  $\Phi_0$ is the superconducting flux quanta,
\begin{equation}
{\rho _s} = \frac{{ \psi  ^2{{e^*}^2}}}{{m^*}},
\end{equation} 
is known as the stiffness, and $e^*$ and $m^*$ are the carriers charge and mass respectively \cite{tinkham2004introduction,schrieffer2018theory,de2018superconductivity}. $ \psi ^2$ is often interpreted as a measure of the superconducting carrier density with a maximum value $ \psi_0 ^2$. When $\nabla \phi=0$ the London equation 
\begin{equation}
{\bf{j}} =  - {\rho _s}{\bf{A}}
\label{JtoA}.
\end{equation} 
is obtained. $\rho_s$ can be expressed in units of length via
\begin{equation}
	{\rho _s} = \frac{1}{\mu_0 \lambda ^2},
	\label{StiffToLambda}
\end{equation}
where $\lambda$ is known as the penetration depth.

The two most important pieces of information on a superconductor (SC) are embedded in Eq.~\ref{JtoAandPhi}. First, $\rho_s$ provides an indication on the ratio between carrier density and effective mass. For example, in high temperature superconductors (HTSC) the
transition temperature $T_c$ is found to be proportional to the stiffness at low temperatures. This finding, known as the Uemura plot, must play a key role in any theory of HTSC \cite{uemura1989universal}. Second, the highest $j$ for which the SC maintains $\nabla \phi=0$ and thus the linear relation of Eq.~\ref{JtoA} holds, sets the critical current $j_c$. $j_c$ also has an interpretation in terms of coherence length via the shortest distance $\xi$ on which $\phi$ can vary by $2\pi$.

However, there is no direct way to measure $\rho_s$. The standard method is to apply magnetic field, to measure the penetration depth of the magnetic induction $\bf{B}$ into a material, and to use Eq.~\ref{StiffToLambda} to determine the stiffness \cite{uemura1989universal,lamhot2015local,morenzoni2002implantation,morenzoni2004nano,prozorov2006magnetic,drachuck2012parallel}. However, magnetic field raises issues one must consider: first, it is essential to take into account the sample shape via the concept of demagnetization factor. This factor is known exactly only for ellipsoidal samples, which are nearly impossible to come by. Second, magnetic fields introduce vortices, which can complicate the interpretation of the penetration depth measurements. Third, all methods have an inherent length scale window. The longest penetration depth that has been measured to the best of our knowledge is $10$~$\mu$m \cite{lamhot2015local,morenzoni2002implantation,morenzoni2004nano,prozorov2006magnetic,drachuck2012parallel}. This is far shorter than a typical sample size. Therefore, there is a temperature range below $T_c$ at which $\lambda > 10$~$\mu$m where the behavior of $\rho_s$ is obscured. For highly anisotropic samples, this range could extend to temperatures well below $T_c$.

Similarly, there is no direct way to measure the critical current density $j_c$. The standard method is to connect leads, and to determine the current at which voltage develops across the sample \cite{zhou2007improved,shay2009interaction,talantsev2014hole,talantsev2015universal}. However, this method could lead to two transitions, first when voltage develops and power, lower than the cooling power, is injected into the sample, second a thermal runaway when the entire sample becomes normal and the voltage grows exponentially \cite{shay2009interaction}. Finally, stiffness and coherence length measurements require different experimental setups.

Here we present in details a new instrument to measure stiffness and coherence length simultaneously, in zero magnetic field and with no leads, based on the London equation (Eq.~\ref{JtoA}). This method determines $\rho_s$ directly without the use of the penetration depth concept. When this equation breaks, and  $\rho_s$ can no longer be determined, it means the critical current has been reached. Consequently, we name the instrument Stiffnessometer. We convert the breaking point of Eq.~\ref{JtoA} to $\xi$ using a mathematical solution of the full Ginzburg-Landau equations in the relevant set up \cite{Gavish2020}. As we explain below, the Stiffnessometer can measure very weak stiffness, which corresponds to $\lambda$ ranging from tens of microns to millimeters. This allows measurements of stiffness closer to the critical temperature $T_c$ than ever before, or measuring the stiffness of very anisotropic systems. Finally, vortices or demagnetization factor are not a problem for the Stiffnessometer since the measurement is done in zero field. The Stiffnessometer was previously used to measure the anisotropy of the stiffness in LSCO $x=0.12$ \cite{kapon2019phase}, but only a brief account to the details of its operation was given.

\section{Experimental setup} \label{setup}

\begin{figure}[tbph]
	\includegraphics[trim=1cm 2cm 1cm 0.4cm, clip=true,width=\columnwidth]{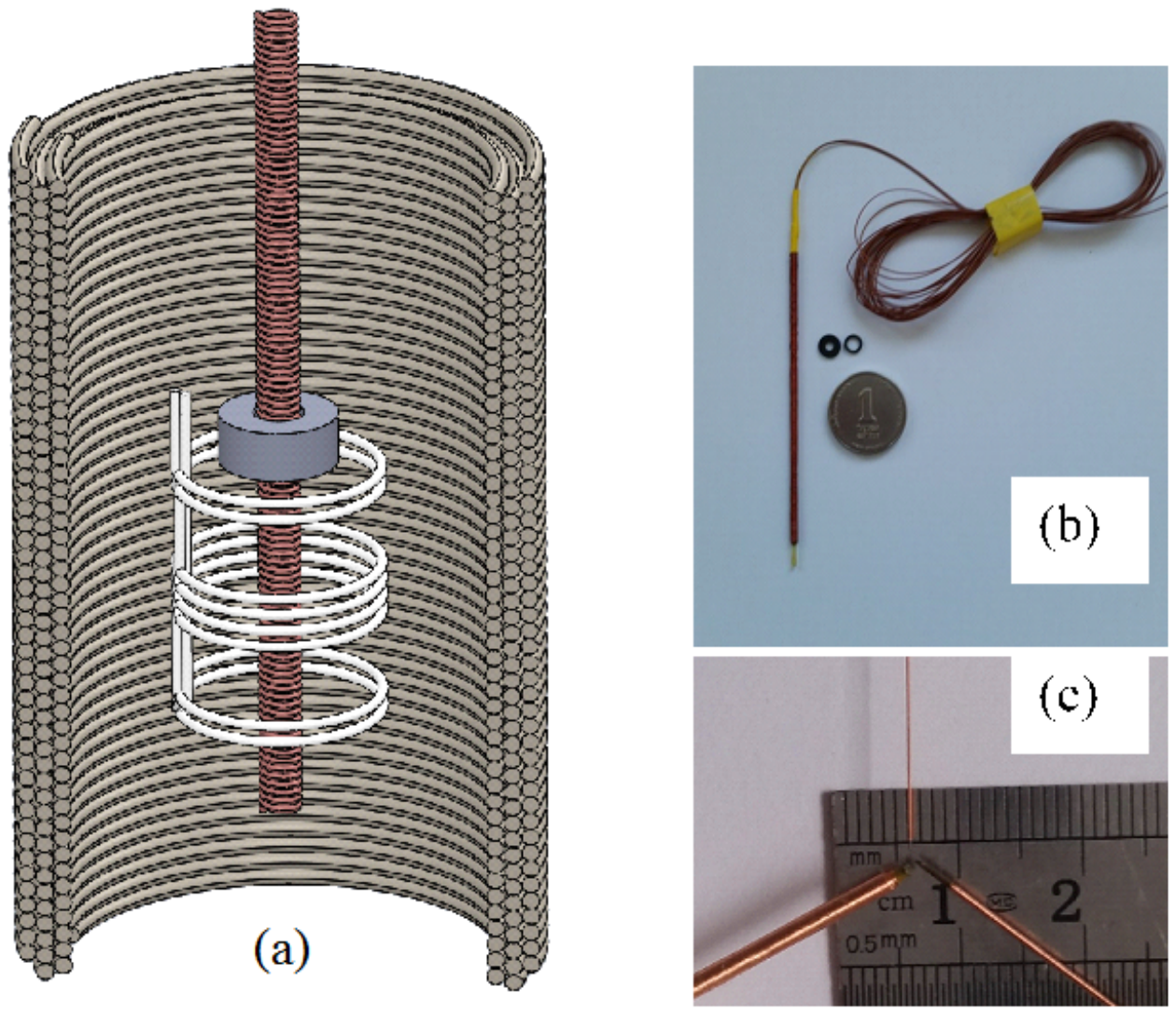}
	\caption{\textbf{Experimental setup.} (a) An illustration of the Stiffnessometer: The superconducting ring is threaded by an inner-coil, placed in the center of a gradiometer, and surrounded by a main-coil that serves as a shim coil. (b) A typical inner-coil, $60$~mm long with $2$~mm outer diameter. Also shown are two La$_{2-x}$Sr$_{x}$CuO$_{4}$ rings with a rectangular cross-section. (c) A zoom-in on other inner-coils with outer diameters ranging from $2.0$~mm to $0.25$~mm, and length of $60$~mm.}
	\label{fig:setup}
\end{figure}

The method is based on the fact that outside an infinitely long coil (defining the ${\bf{\hat z }}$ direction), the magnetic field is zero while the vector potential is finite. This vector potential is tangential and points in the ${\bf{\hat \varphi }}$ direction. When such an inner-coil is placed in the center of a SC ring, the vector potential leads to a current density in the ring according to Eq.~\ref{JtoAandPhi}. This current flows around the ring and generates a magnetic moment, which is detected by moving the ring and the inner-coil rigidly relative to a pickup-loop. The concept of the measurement is depicted in Fig.~\ref{fig:setup}(a). A typical inner-coil and two superconducting rings of the cuprate SC La$_{2-x}$Sr$_{x}$CuO$_{4}$ (LSCO) are shown in Fig.~\ref{fig:setup}(b). In Fig.~\ref{fig:setup}(c) we present a zoom-in on three different coils with outer diameters of $2$, $0.8$, and $0.25$~mm. They have 2 to 16 layers of wires with thickness between 10 and 100 $\mu$m, and their length is $60$~mm. Our Stiffnessometer is an add-on to a Cryogenic SQUID and to a quantum design MPMS3 magnetometers.

Both magnetometers use a second order gradiometer, rather than a single pickup-loop. The gradiometer is made of three winding groups. The outer two are constructed from two loops each, wound clockwise, and the inner group is made of 4 loops, wound anticlockwise. This is also demonstrated in Fig~\ref{fig:setup}(a). The gradiometer ensures that a magnetic moment generates voltage only when it is in the vicinity of the gradiometer center. Also, any field uniform in space gives zero signal even if it drifts in time. The gradiometer is connected to a superconducting quantum interference device (SQUID). The output voltage $V$ of the device is proportional to the difference between flux threading the different loops of the gradiometer.

\begin{figure}[tbph]
	\includegraphics[trim=0cm 1cm 1cm 0.4cm, clip=true,width=\columnwidth]{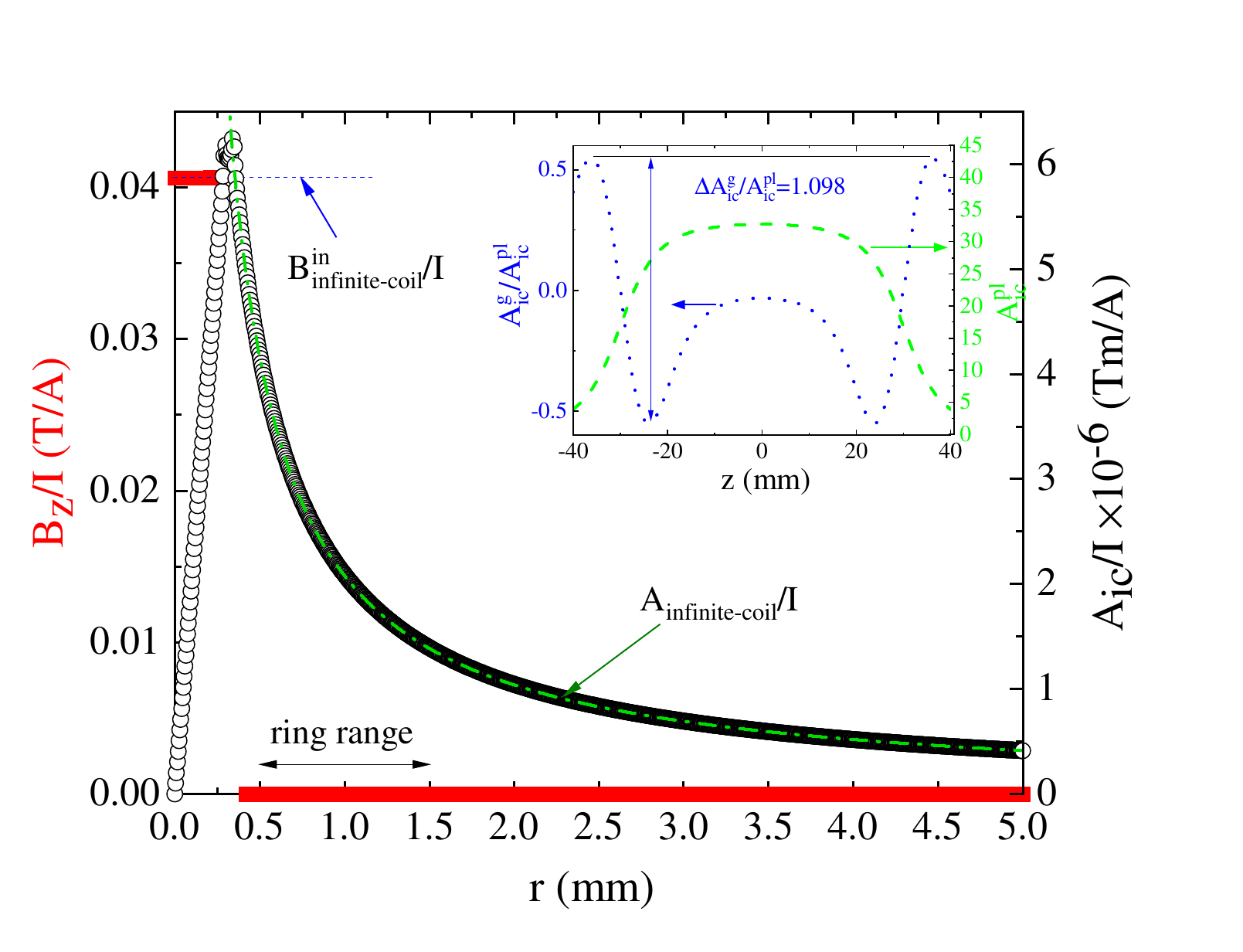}
	\caption{\textbf{Vector potential and magnetic field profile.} Numerical calculation of the vector potential and magnetic field per current at $z=0$ for the inner-coil used in this study. The coil parameters are: length $l=60$~mm, inner diameter $=0.54$~mm, outer diameter $=0.8$~mm, 2 layers and 1940 turns. The ring position relative to the inner-coil center is demonstrated by the double arrows. The vector potential is very well approximated by an infinite coil over the range of the ring as the dashed-doted green line demonstrates. Inset: $A_{ic}^{pl}$ and $A_{ic}^g/A_{ic}^{pl}$ as a function of z, as explained in the main text.}
	\label{BandAvsr}
\end{figure}

The vector potential outside of an infinitely long coil is given by
\begin{equation}
{\textbf{A}_{ic}} = \frac{{{\Phi _{ic}}}}{{2\pi r}}{\bf{\hat \varphi }},
\label{AicDef}
\end{equation} 
where $r$ is the distance from the center of the coil, and ${\Phi _{ic}}$ is the flux produced by the inner-coil. To check the validity of this expression in our case we calculated numerically the magnetic field $B_z$ and vector potential $A_{\varphi}$ (in the Coulomb gauge) produced by the inner-coil as a function of $r$ and $z$. This coil is $60$~mm long, has inner diameter (I.D.) of $0.54$~mm, outer diameter (O.D.) of $0.8$~mm, $2$ layers, and $1940$ turns in total. The measured LSCO ring has an I.D. $1.0$~mm, O.D. $2.5$~mm, and height ($h$) of $1.0$~mm. Fig.~\ref{BandAvsr} shows the result of the calculations. The  approximation of an infinite coil, presented by the dashed-doted green line, is perfect for our ring size and even for much larger rings. The calculation also shows that the strongest field just outside of the inner-coil is $10^{4}$ times smaller than the field at its center.

The sample is grown using Optical Floating Zone Furnace. It is oriented using x-ray Laue camera and cut to plates and then into a ring shape using ELAS Master femto-second laser cutter. The ring's plane is the CuO$_2$ plane of the sample. After cutting, the sample is annealed at 850 C$^\circ$ for 120 hours in argon atmosphere.

The measurements are done in two different detection methods. (I) DC scan mode, where we record the SQUID's output voltage $V(z)$ while the relative distance between the gradiometer and the ring changes when the ring and inner-coil move. The DC mode allows detection of the contribution from the inner-coil as well, since the entire coil can be pulled out of the gradiometer. Our gradiometer detects magnetic moments within a range of $15$~mm on each side of its center. This sets the length of our inner-coils.  When measuring over a wide temperature range, detection of the inner-coil contribution is important in order to determine the flux it generates at each temperature. (II) VSM mode, where the ring vibrates around the center of the gradiometer. In this mode the coil does not contribute to the signal. The VSM mode is fast and allows fine temperature scans without the need to achieve temperature stability at each measuring point.

There is a risk that field generated in the inner-coil leaks since no coil is infinitely long or perfect. To overcome this leak, a main coil, also shown in Fig.~\ref{fig:setup}(a), acts as a shim to cancel the field on the ring when it is at the gradiometer center. In the Cryogenic SQUID the main-coil has a field resolution of $0.1$~$\mu$T. The Ultra-low field (ULF) capability of MPMS3 allows for field cancellation down to $0.3$~$\mu$T. Therefore, we can keep the field on the ring as low as $0.1$~$\mu$T when needed.

\begin{figure}[tbph]
	\includegraphics[trim=1.0cm 1cm 1cm 0.4cm, clip=true,width=\columnwidth]{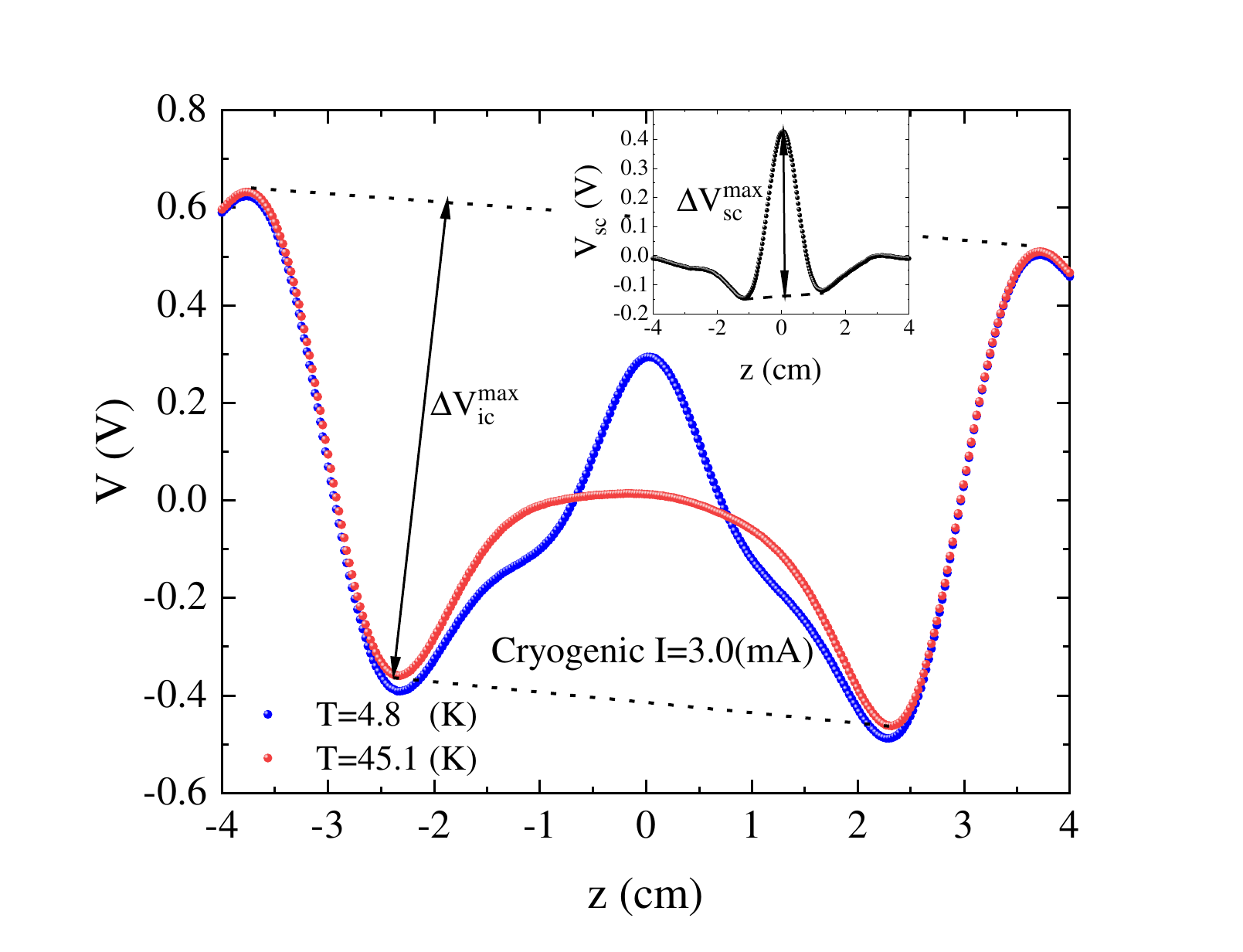}
	\caption{\textbf{Raw Data.} SQUID signal for a LSCO $x=0.17$ ring at high temperature, when the ring is not superconducting, and at low temperature when the ring is superconducting. The inset shows the difference between these measurements}
	\label{RawData}
\end{figure} 

The measurements can be done in two different procedures: One, is zero gauge field cooling (ZGFC) in which we cool the ring to a temperature below $T_c$, turn on the current in the inner-coil $I$ when the ring is superconducting, and measure while warming. In this procedure, the SC minimizes its free energy by setting $ \nabla \phi =0 $ in Eq.~\ref{JtoAandPhi}. This value of $ \nabla \phi $ does not change as $\textbf{A}$ is turned on, as long as the current in the coil is below some critical value (as explained later). In this case Eq.~\ref{JtoA} holds throughout the measurements. The other procedure is gauge field cooling (GFC) in which we turn on the current in the inner-coil at a temperature above $T_c$, cool the inner-coil and ring below $T_c$, and turn the current off. To minimize its free energy the SC sets $ \nabla \phi $ in Eq.~\ref{JtoAandPhi} such that $j$ is as close to zero as possible. When $\bf{A}$ is turned off, $\nabla \phi$ does not change and plays the role of $\bf{A}$ in the ZGFC procedure.

To better appreciate why $\nabla \phi =0$, even when $A$ is ramped, one can view $\phi$ as the phase of an in-plane arrow. Cooling at $A=0$ sets all the arrows pointing in the same direction. Since the phase is quantized, to change $\phi$ means a twist of all arrows in a closed loop, such that the phase between the first arrow and last one in the loop changes by $2\pi$. This would lead to a discontinuity in the phase value, a procedure that costs energy, and generates instantaneous voltage according to the Josephson equation $\frac{\hbar }{{{e^*}}}\frac{{\partial \phi }}{{\partial t}}$. A nice analog is a ferromagnetic ring with the spins pointing in the same direction. Rotating the last spin with respect to the first one by $2\pi$ requires to break a bond. This procedure is not energetically favorable for a ferromagnet (or the SC ring). Therefore, ramping $A$ leaves all arrows pointing in the same direction and $\nabla \phi=0$, until $A$ exceeds a critical value. At this point, the current is too high and it is worthwhile for the SC to ``break a bond" and reduce the current.

\begin{figure*}[h!t]
	\begin{center}
		\includegraphics[trim=0cm 1cm 0.5cm 0.4cm, clip=true,width=15cm]{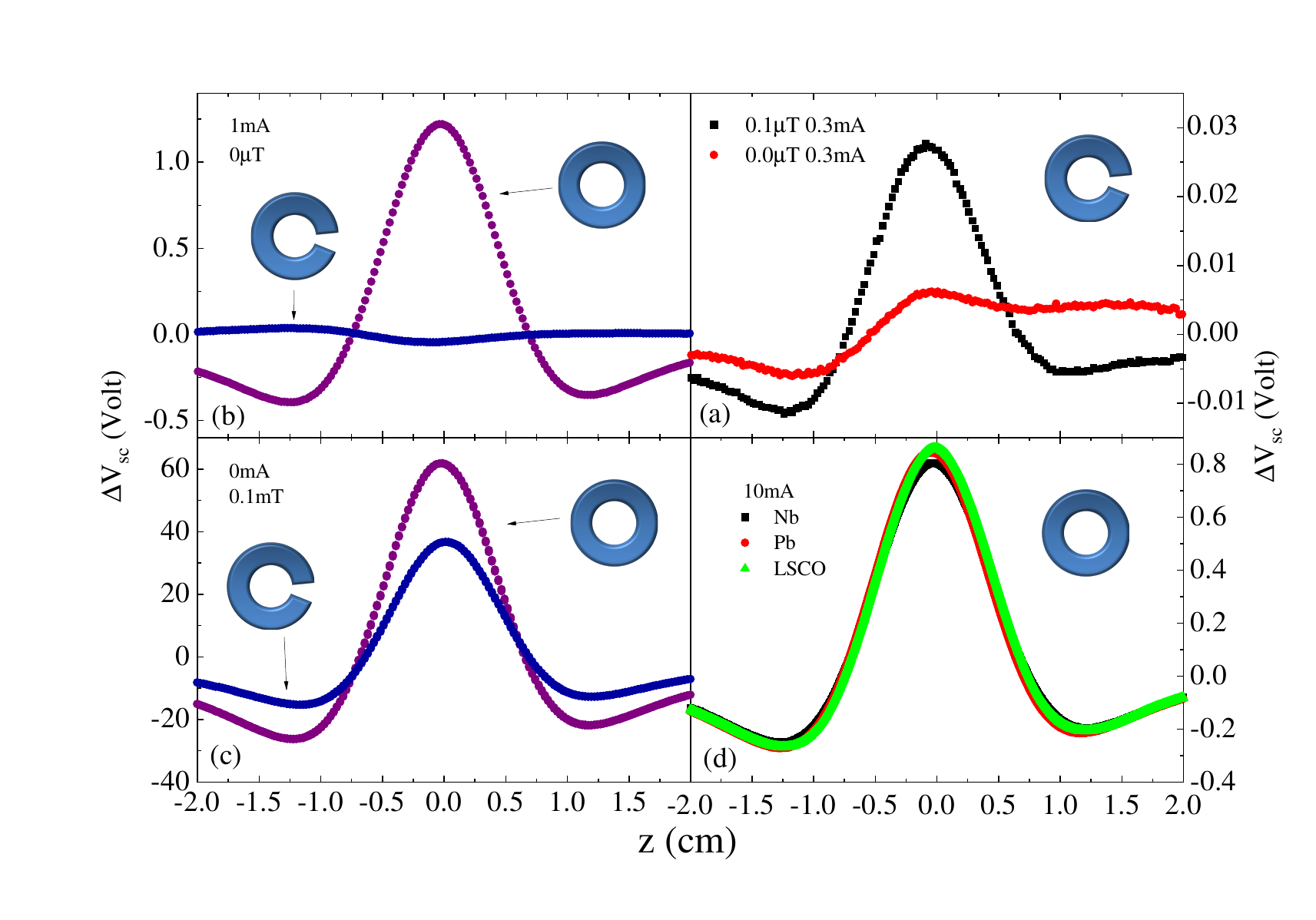}
	\end{center}
	
	\caption{\textbf{Experimental tests.} (a) The signal with a current of $0.3$~mA in the inner-coil and $0.1$~$\mu$T fields demonstrating the quality of the field cancelling procedure. (b) The SQUID signal for an open and closed rings when the field is zero and the vector potential is finite. (c) A test experiment: the SQUID signal for an open and closed rings when the vector potential is zero but the field is finite. (d) Demonstrating that when $\lambda$ is much smaller than the sample size the signal is material independent.}
	\label{fig:tests}
\end{figure*}

A typical DC mode measurement is demonstrated in Fig.~\ref{RawData}. The red symbols represent the signal when the entire inner-coil has moved through the pickup-coil at $T>T_c$. Before the lower-end of the inner-coil has reached the gradiometer, the flux through it is zero. During the time the lower-end of the inner-coil transverse the gradiometer its contribution to the total flux changes from zero to positive to negative and back to zero. The upper-end of the inner-coil has the opposite effect; its contribution to the flux goes from zero to negative to positive and back to zero. But there is a time (or distance) delay between the lower-end and upper-end contributions, leading to the observed signal. A linear drift of the voltage can be easily evaluated as demonstrated by the dotted lines. We define the inner-coil maximum voltage difference $\Delta V_{ic}^{max}$ as demonstrated in Fig.~\ref{RawData}.

At $T<T_c$ the ring adds its own signal, as shown in Fig.~\ref{RawData} by blue symbols. The ring produces current that generates opposite flux to the one in the inner-coil. The ring signal is concentrated on a narrower range on the $z$ axis. By subtracting the high temperature measurement from the low temperature one, it is possible to obtain the signal from the ring alone $V_{sc}$ as demonstrated in the inset of Fig.~\ref{RawData}. We define the maximum ring voltage difference $\Delta V_{sc}^{max}$ as shown in the inset. The ratio $\Delta V_{sc}^{max}/\Delta V_{ic}^{max}$ stores the information on the stiffness, as will be discussed in the Data Analysis Sec.~\ref{DataAnalysis}.

\section{Tests} \label{sec:tests}

To ensure that our signal is not due to leakage of magnetic field from the inner-coil or any other field source, we perform three tests. In the first one we apply current in the inner-coil, measure the field leakage at the ring position using an open ring and cancel it using the main coil. Then we increase the field by only $0.1$~$\mu$T. The measurements before and after the field increase are depicted in Fig.~\ref{fig:tests}(a). They indicate that we can cancel the field in the ring position to better than $0.1$~$\mu$T. Clearly in zero field there is no signal. In the second test we measure the stiffness (zero field and applied current in the inner-coil) of closed and open rings, which are otherwise identical in size. The results are shown in Fig.~\ref{fig:tests}(b). The signal from a closed ring is much bigger than the background from an open one. In Fig.~\ref{fig:tests}(c) we repeat this measurement with an applied field in the main coil of $0.1$~mT, and no current in the inner-coil. In this case both open and closed rings give strong and similar signals. The difference between the two signals is consistent with the missing mass in the open ring. These tests confirm that the field leakage is not relevant to our stiffness measurement. Our ability to determine small stiffness depends on how well we can cancel the field at the ring position.

Another important test of the Stiffnessometer comes from comparing the signal from rings of exactly the same dimensions, but made from different materials. At temperatures well below $T_c$ the stiffness is expected to be strong, namely, the penetration depth should be much shorter than all the ring dimensions. In this case, as the current is turned on, and flux in the inner-coil ${{\Phi _{ic}}}$ changes, an electric field is generated in the SC ring $E_{sc}$ according to 
\[{E_{sc}} = \frac{1}{{2\pi r}}\frac{{\partial {\Phi _{ic}}}}{{\partial t}} =  -\frac{{\partial {A_{sc}}}}{{\partial t}}\]
where $A_{sc}$ is the vector potential of the ring. This leads to 
\[{\Phi _{sc}} = 2\pi r{A_{sc}} =  - {\Phi _{ic}},\] 
where $\Phi _{sc}$ is the flux generated by the SC ring at its center. In other words, when $\lambda$ is short compared to the ring dimensions, the SC produces flux which exactly cancels the applied flux through it, regardless of the material used. Therefore, all materials should produce the same signal. This is demonstrated in Fig.~\ref{fig:tests}(d) for Niobium (Nb), Lead (Pb) and LSCO. They all have the same $\Delta V_{sc}$.

\section{Measurements} \label{Measurements}

In this section we present mainly Stiffnessometer raw data out of which we are able to extract $\rho_s$, $\xi$, and $j_c$ as a function of temperature in favorable conditions.

\subsection{Stiffness and its temperature dependence}

\begin{figure}[tbph]
	\includegraphics[trim=1cm 0.5cm 1cm 0.4cm, clip=true,width=\columnwidth]{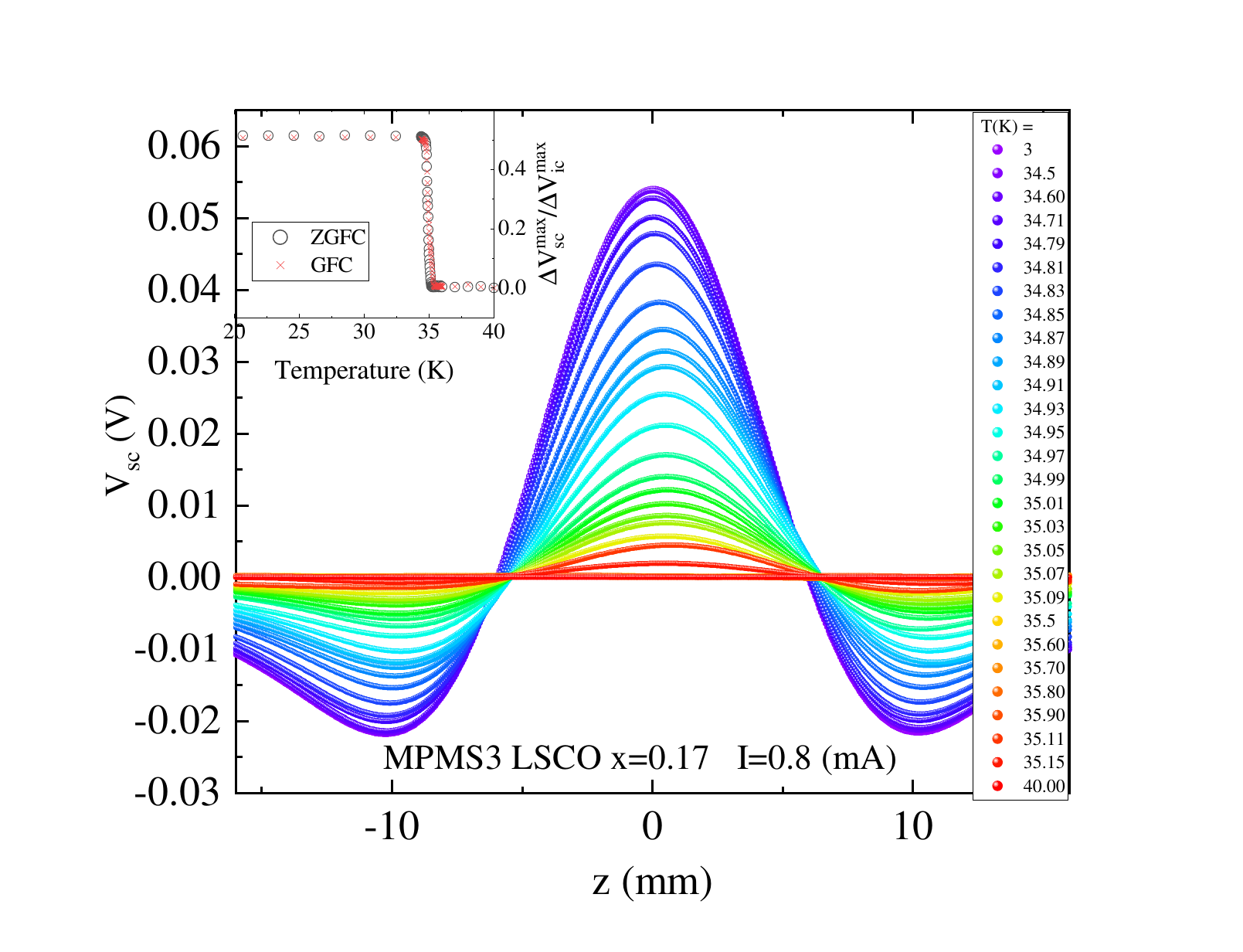}
	\caption{\textbf{Temperature dependence.} The SQUID signal $V_{sc}$ for a La$_{2-x}$Sr$_{x}$CuO$_{4}$ $x=0.17$ ring with the CuO$_2$ planes perpendicular to the ring symmetry axis, at different temperatures. The inset show  $\Delta V_{sc}^{max}/\Delta V_{ic}^{max} $ in the ZGFC and GFC procedures as a function of temperature}
	\label{TDep}
\end{figure}

In Fig.~\ref{TDep} we present the Stiffnessometer signal evolution with temperature for the LSCO $x=0.17$ ring as measured by the DC mode and ZGFC procedure with $I=0.8$~mA. At temperatures between $3.0$~K and $34.7$~K there is no change in the signal. But, between $34.7$~K and $T_c=35.53$~K the signal diminishes rapidly, as expected. The inset of Fig.~\ref{TDep} shows $\Delta V_{sc}^{max}/\Delta V_{ic}^{max}$ from both ZGFC and GFC measurement protocols. There is no difference between the two strategies.

\subsection{Critical current and its temperature dependence} \label{Criticalcurrent}

\begin{figure}[tbph]
	\includegraphics[trim=0cm 0cm 0cm 0cm, clip=true,width=\columnwidth]{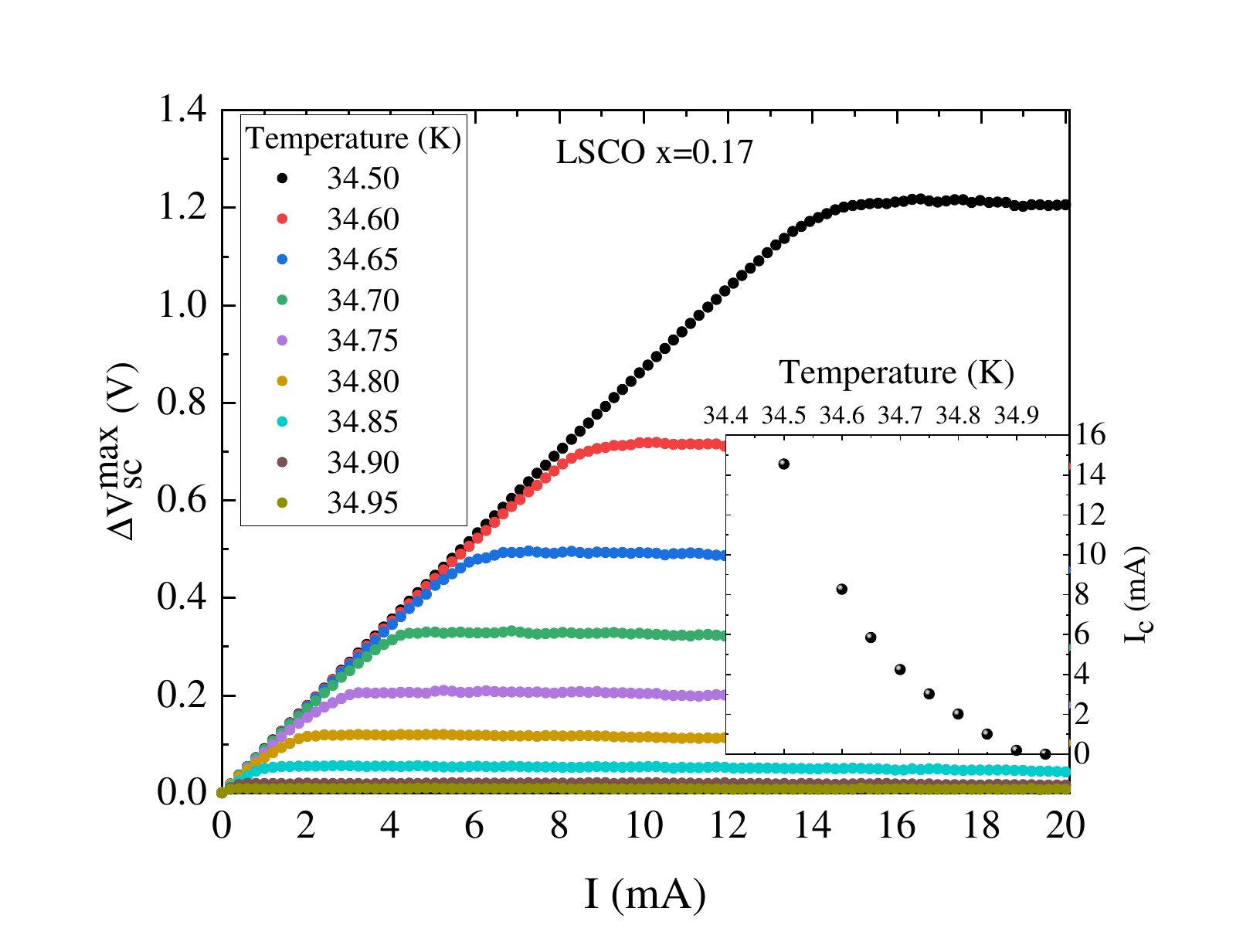}
	\caption{\textbf{Critical currents.} The SC ring signal $\Delta V_{sc}^{max}$ as a function of applied current in the inner-coil $I$, for different temperatures approaching $T_c$. The inset shows critical current $I_c$, where the signal becomes current independent, as a function of temperature}
	\label{fig:CriticalCurrent}
\end{figure}

The Stiffnessometer can also be used to measure critical currents. This is depicted in Fig.~\ref{fig:CriticalCurrent} for the LSCO ring at various temperatures. The signal from the ring $\Delta V_{sc}^{max}$ grows linearly with $I$ at each $T$, but abruptly becomes $I$ independent at a critical current $I_c (T)$, presented in the inset. It means that the SC can generate only a finite amount of opposing flux. Therefore, we are detecting $j_c$ of the SC.

As $I$ exceeds $I_c$, vortices start to flow into the center of the ring, so that $j$ in the ring never exceeds $j_c$. In other words, once the critical current in the sample is crossed, $\nabla \phi$ is no longer zero and becomes $\nabla \phi =m/r$ with $m\ne 0$. The SC selects $m$ such that $j$ is fixed. Therefore, for $I > I_c $, the current in the ring and $\Delta V_{sc}^{max}$ are fixed.

\section{Data Analysis} \label{DataAnalysis}

Analyzing the Stiffnessometer signal is done in steps: (A) we consider a single pickup-loop and then a gradiometer. (B) The order parameter magnitude $\left| \Psi  \right|$ is taken to be constant in space and the stiffness is weak. Weak stiffness means that the vector potential on the ring is only due to the applied current. The vector potential generated by the internal current of the ring is ignored. This approximation is valid when the ring's current density is smaller than $j_c$ and the penetration length is longer than the sample dimensions. The weak stiffness analysis is analytical, and valid close (but not too close) to $T_c$. (C) The order parameter is still assumed to be constant in space but now the stiffness is strong. In this case, the self vector potential is taken into account. This leads to a partial differential equation (PDE), which we solve numerically with relatively simple means. (D) A full solution of the coupled Ginzburg-Landau equations allowing for both $\left| \Psi  \right|$ and $A$ to be space dependent. This level of analysis is required only when the SC is nearly destroyed by the internal currents, and it is good for extracting $j_c$ and $\xi$. This level of analysis, for the case of a very tall hollow cylinder, is covered in Ref.~\cite{Gavish2020}. Consequently, at present we can only place limits on $j_c$ and $\xi$.

\subsection{Single pickup-loop and gradiometer}

Had we used a single pickup-loop, the voltage would have been proportional to the flux threading it $\Phi=2\pi R_{pl} A(R_{pl})$, where $R_{pl}=13$~mm and $R_{pl}=8.5$~mm for the Cryogenic and MPMS2 pickup-loop radii respectively. Above $T_c$, maximum voltage is achieved when the pickup-loop is at the center of the inner-coil so that ${V_{ic}^{max}} = k 2 \pi R_{pl}A_{ic}(R_{pl},z=0)$ where $k$ is a proportionality constant. Similarly, a ring at the center of and parallel with a pickup-loop would generate a maximum voltage proportional to its own flux, $V_{sc}^{max}=k 2 \pi R_{pl}A_{sc}(R_{pl},z=0)$ where $A_{sc}$ is the vector potential generated by the ring. Therefore, 
\begin{equation}
\frac{V_{sc}^{max}}{V_{ic}^{max}}=\frac{A_{sc}(R_{pl},z=0)}{A_{ic}(R_{pl},z=0)}.
\label{SingleLoop}
\end{equation}

Next, we convert between the signal detected by a gradiometer to the signal that would have been detected by a single pickup-loop. We find a conversion factor, $G$, from the vector potential evaluated on a single pickup-loop $A^{pl}$ to the differences in the vector potential generated by the gradiometer $\Delta A^g$. This has to be done for both the ring and the inner-coil.
The vector potential of a ring with magnetic moment $\mathfrak{m}$ on the pickup-loop depends on the moment's height $z$ from the plane of the loop according to $A=2\pi \mathfrak{m} R^2_{pl} /(R_{pl}^2+z^2)^\frac{3}{2}$. Therefore, for a ring and our gradiometer
\begin{multline}
\frac{A^g_{sc}(R_{pl},z)}{A^{pl}_{sc}(R_{pl},z=0)}=\frac{-2R_{pl}^3}{(R_{pl}^2+(z+\Delta z_{pl})^2)^\frac{3}{2}}+\frac{4R_{pl}^3}{(R_{pl}^2+z^2)^\frac{3}{2}}+\\ +\frac{-2R_{pl}^3}{(R_{pl}^2+(z-\Delta z_{pl})^2)^\frac{3}{2}},
\end{multline}
where $\Delta z_{pl}=7.0$~mm and $\Delta z_{pl}=8.0$~mm is the separation between the different groups of gradiometer windings for Cryogenic and MPMS3 magnetometers receptively. The difference between the maximum and minimum of this function is $\Delta A_{sc}^g/A_{R}^{pl}=1.70$ and $3.37$, again respectively, are the conversion factor for the ring.

To convert from $A_{ic}^{pl}$ to $\Delta A_{ic}^g$ we plot by the green line in the inset of Fig.~\ref{BandAvsr} the vector potential generated by our coil at $R_{pl}$ as a function of $z$, $A_{ic}^{pl}(z)$. The plot is specific for $\Delta z_{pl}=7.0$~mm. The function
\begin{equation}
\small
\frac{A^g_{ic}(z)}{A_{ic}^{pl}}=\frac{-2A_{ic}^{pl}(z+\Delta z_{pl}) + 4A_{ic}^{pl}(z) - 2A_{ic}^{pl}(z-\Delta z_{pl})}{ A_{ic}^{pl}(0)}
\end{equation}
is also plotted in the inset by the blue line. The difference between the maximum and minimum of this function is the conversion factor for the inner-coil. We find numerically that $\Delta A^g_{ic}/A_{ic}^{pl}=0.47$. Thus
\begin{equation}
\frac{\Delta V_{sc}^{max}}{\Delta V_{ic}^{max}}=G\frac{A_{sc}^{pl}}{A_{ic}^{pl}}
\label{eq:Vratio_to_Aratio}
\end{equation}
with $G=3.62$ and $3.07$ for Cryogenic and MPMS3 magnetometers respectively. By measuring ${\Delta V_{sc}^{max}}/{\Delta V_{ic}^{max}}$ one can predict the expected vector potential ratio between the coil and the ring at the pickup loop position. As we show below, $G$ could also be calibrated experimentally.

As for the VSM method, the magnetic moment $\mathfrak{m}$ of the ring and $A_{sc}^{pl}$ are related by
\begin{equation}
\frac{\mathfrak{m}}{\Delta V_{ic}^{max}}=F\frac{A_{sc}^{pl}}{A_{ic}^{pl}}
\label{eq:m_to_A}
\end{equation}
where $F$ is a calibration factor. In the GFC procedure $\Delta V_{ic}^{max}$ is measured before the coil current is turned off. $F$ is determined by measuring $\mathfrak{m}$, and calculating ${A_{sc}^{pl}}/{A_{ic}^{pl}}$ in conditions that are not sensitive to the stiffness, as we do below.

\subsection{Weak stiffness, $\left| \psi(\bf{r})  \right|=\psi_0 $} \label{Weak}

The current from each ring element is $j(r)hdr$ where $h$ is the ring height and $dr$ is a ring element width. Using the London equation, the magnetic moment generated by each ring element is $d\mathfrak{m} =  \frac{{r{\rho _s}{\Phi _{ic}}h}}{{2c}}dr$. Integrating from the inner to the outer radii yields the total moment  of the ring $\mathfrak{m} = \frac{{{\rho _s}{\Phi _{ic}}h}}{{4c}}(r_{out}^2 - r_{in}^2)$, and 
\begin{equation}
A_{sc} = \frac{\mathfrak{m}}{r^2}
\label{DVtoMoment}
\end{equation}
Using Eq.~\ref{StiffToLambda}, the penetration depth is given by
\begin{equation}
{\lambda ^2} =  \frac{{h(r_{out}^2 - r_{in}^2)}}{8{R_{pl}}} \frac{A_{ic}(R_{pl})}{A_{sc}(R_{pl})}. 
\label{WeakStiff}
\end{equation}
Since all the dimensions of the ring and pickup-loop are on the order of $ 1$~mm, and we can measure voltage ratios to better than 5\%, we can measure $\lambda$ on the order of 1~mm.

\subsection{Strong stiffness, $\left| \psi(\bf{r})  \right|=\psi_0 $} \label{Strong}

\begin{figure}[tbph]
	\includegraphics[trim=0.5cm 1cm 1cm 0.4cm, clip=true,width=\columnwidth]{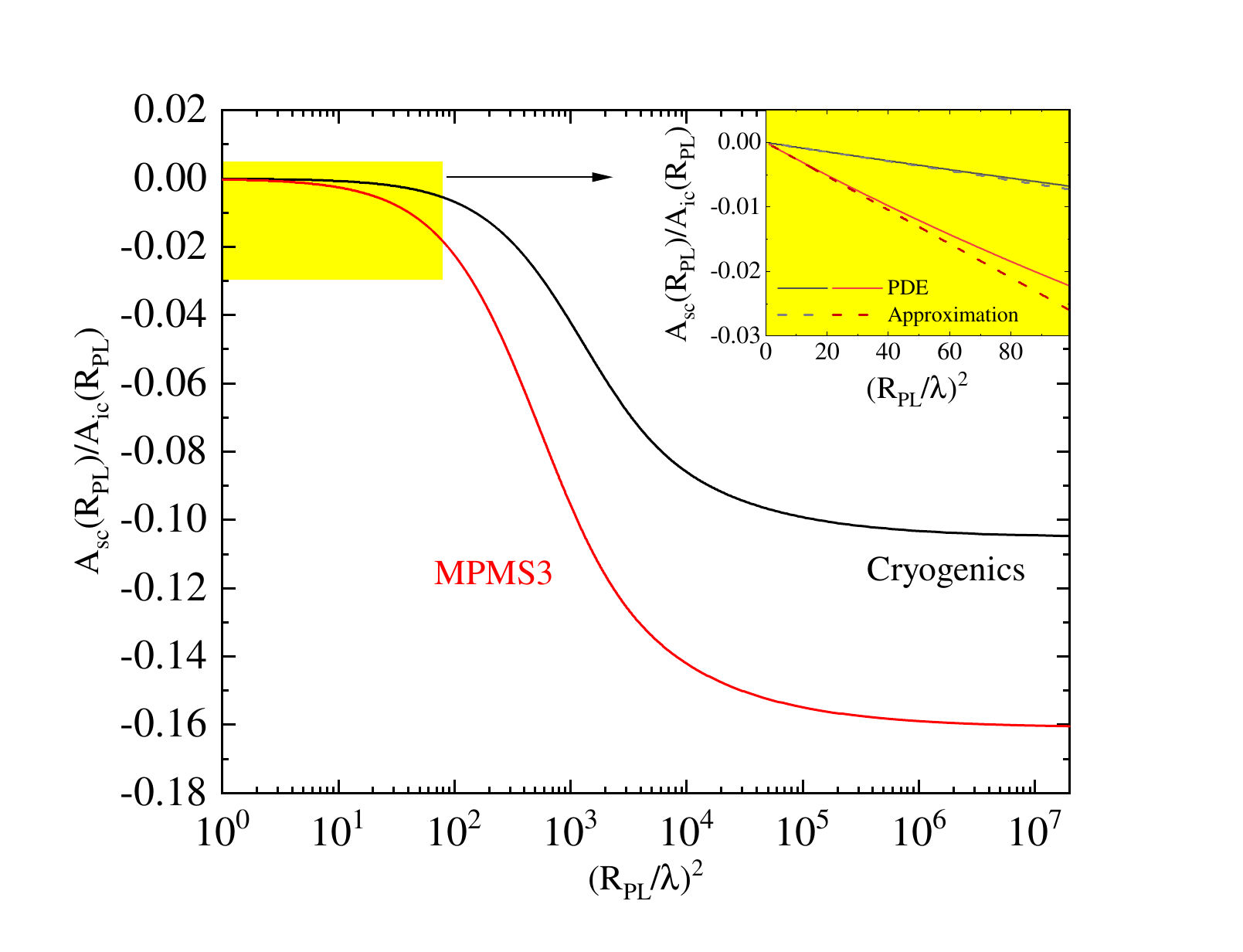}
	\caption{\textbf{Solution of the Stiffnessometer PDE.} A semi-log plot of the solution of Eq.~\ref{KGK} evaluated at the pickup coil radius, for different values of $(R_{pl}/\lambda)^2 $. The inset shows the behavior for large $\lambda$. The solid line is given by Eq.~\ref{WeakStiff} }
	\label{PDE}
\end{figure}

In the strong stiffness case, the total vector potential experienced by the ring ${{\bf{A}}_t}$ is the sum of ${{\bf{A}}_{ic}}$ and ${\bf{A}}_{sc}$. Using Faraday's and London's equations, with ${\bf{B}} = \nabla  \times {\bf{A}}$, and the transformation $ \psi({\bf{r}})  / \psi_0   \to  \psi({\bf{r}}) $ one finds that
\begin{equation}
\nabla ^2 {\bf{A}}_{sc} = \frac{\psi^2({\bf{r}})}{{{\lambda ^2}}}\left( {\frac{{{\Phi _{ic} }}}{{2\pi r}}{\bf{\hat \varphi }} + {{\bf{A}}_{sc}}} \right)
\label{eq:PDE}
\end{equation}
where $\psi({\bf r})=1$ inside the SC and zero outside. The Coulomb gauge is built into Eq.~\ref{PDE} inside the ring since for any vector field $\bf{F}$, $\nabla  \cdot \nabla  \times {\bf{F}} = 0$. Outside of the ring, this gauge has to be imposed separately. In cylindrical coordinates, ${\bf{A}}_{sc} = A(z,r){\hat \varphi }$, and with the coordinate transformation
\begin{equation}
{\bf{r}}/R_{pl}  \to {\bf{r}}, {\bf{A}}_{sc}/{A}_{ic}(R_{pl}) \to {\bf{A}}, \lambda/R_{pl} \to \lambda
\end{equation}
the equation in the ring becomes
\begin{equation}
\frac{{{\partial^2}A}}{{\partial{z^2}}}+\frac{{{\partial^2}A}}{{\partial{r^2}}} + \frac{1}{r}\frac{{\partial A}}{{\partial r}} - \frac{A}{{{r^2}}}=  \frac{ \psi^2({\bf{r}}) }{\lambda^2}\left(A + \frac{1}{r}\right) 
\label{KGK}
\end{equation}
with $r$, $z$, and $\lambda$ are in units of $R_{pl}$, and $A$ is in units of $A_{ic}(R_{pl})$. The solution of this equation, evaluated at $R_{pl}$, is the quantity one would measure with a single pickup-loop as indicated in Eq.~\ref{SingleLoop}.

\begin{figure}[tbph]
	\includegraphics[trim=0.0cm 3.0cm 2.0cm 2.0cm, clip=true,width=9cm,keepaspectratio]{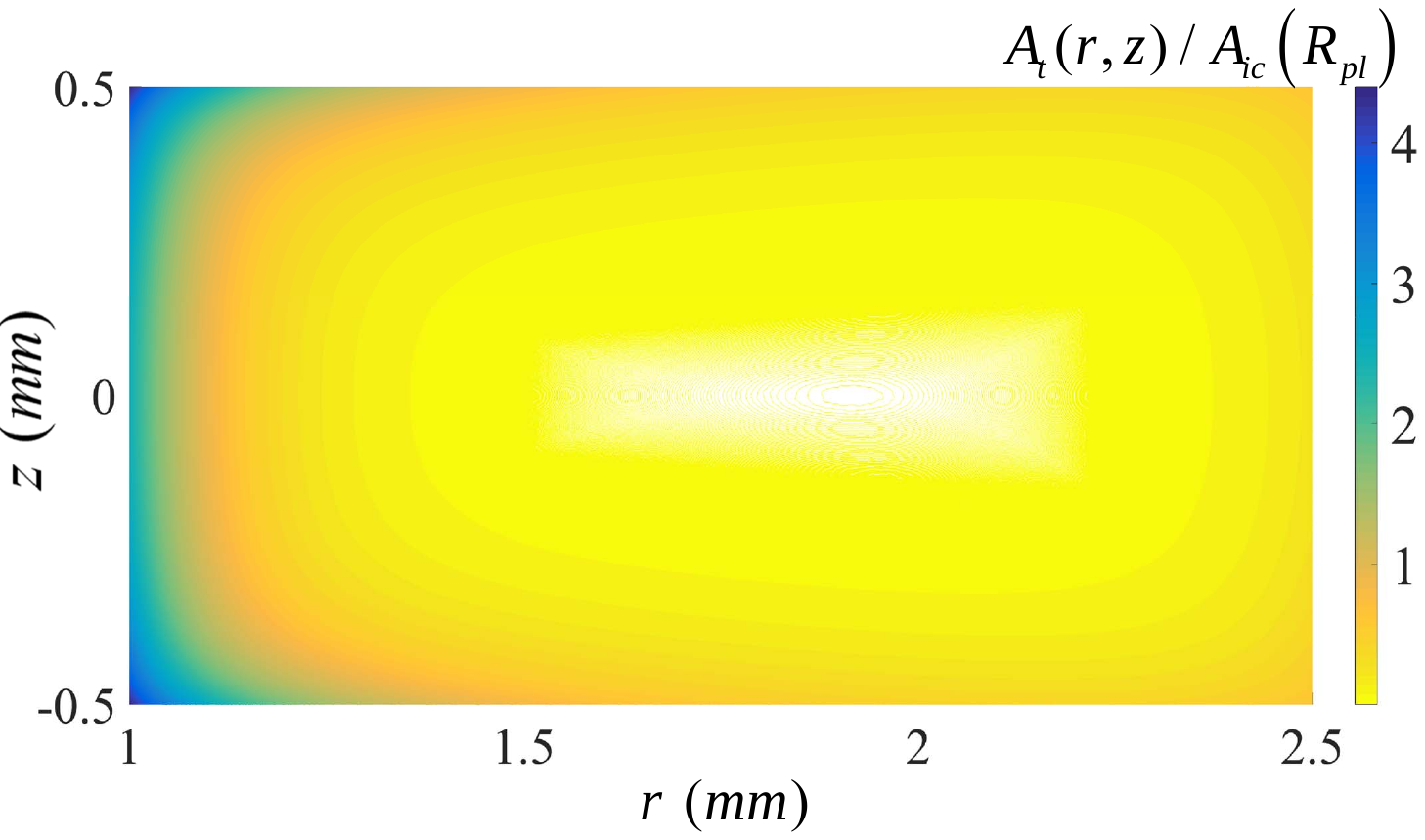}
	\caption{\textbf{$A_{t}$ distribution inside the ring.} The total vector potential obtained from the solution of Eq.~\ref{KGK} and the vector potential of the inner-coil $A_{ic}$, as a function of $r$ and $z$ for $\lambda/R_{pl}=0.1/13$, $r_{in}=1$~mm, $r_{out}=2.5$~mm, $h=1$~mm. }
	\label{Atot}
\end{figure} 

We solved Eq.~\ref{KGK} for different $\lambda$ values and our LSCO ring parameters with both the Comsol 5.2a and FreeFem \cite{freefem} softwares. We used finite elements in a box $[-L_z,L_z]\times[0,L_r]$ where $L_z=L_r=8$. Dirichlet boundary conditions are imposed at $z=\pm L_z$, $r=0$, and $r=L_r$. Maximal mesh spacing is set to be $h=0.01$ in the ring and its immediate vicinity, and $h=0.25$ elsewhere. The total vector potential $A_{t}$ for $\lambda/R_{pl}=0.1/13$, and for all values of $r$ and $z$ in the ring cross section is presented in Fig.~\ref{Atot}. Clearly, the vector potential, hence the current, is strongest close to the inner radius of the ring. They decay towards the center of the ring. The solutions at $r=1$ and $z=0$ and our ring parameters, for a range of $\lambda$ values, and different magnetometers, are presented in Fig.~\ref{PDE} on a semi-log plot. The inset is a zoom-in on the long $\lambda$ region emphasized by a rectangle. The solid line represents Eq.~\ref{WeakStiff} again with our LSCO ring parameters. There is a good agreement between the PDE solution at long $\lambda$ and the weak-stiffness approximation.

In Fig.~\ref{PDE} we see that when the penetration depth is very short, $A_{sc}^{pl}/A_{ic}^{pl}=-0.16$ for the MPMS3. Multiplying the absolute value of this number by the MPMS3 $G=3.07$ we expect a saturation value of $\Delta V_{sc}^{max}/\Delta V_{ic}^{max}=0.49$. The measured value, however, is $0.516$ as seen in the inset of Fig.~\ref{TDep}. The calculated and experimental $G$ factors are somewhat different. The experimental ``G factor" is determined by dividing the measured saturation voltage ratios by the numerical saturation value. For the presented data of LSCO $x=0.17$ this yields $G=3.22$.

\subsection{Ginzburg-Landau} \label{GinzburgLandau}

When the current $j$ somewhere in the SC is strong enough to destroy superconductivity, $\psi$ becomes space dependent even inside the SC. One has to solve two Ginzburg-Landau equations simultaneously. Consider a hollow long cylinder. Using the transformation: $2\pi R_{pl} A_{sc}/\Phi_0 \to A_{sc}$ and normalizing all lengths by $R_{pl}$ these equations are given by

\begin{equation}\label{eq:Asc}
\frac{{{\partial^2}A_{sc}}}{{\partial{r^2}}} + \frac{1}{r}\frac{{\partial A_{sc}}}{{\partial r}} - \frac{A_{sc}}{{{r^2}}}=\frac{\psi^2(r)}{\lambda^2}\left(A_{\rm sc}+\frac{J}{r}\right),
\end{equation}
and
\begin{equation}\label{eq:GL2}
\xi^2\left(\frac{{{\partial^2}\psi}}{{\partial{r^2}}} + \frac{1}{r}\frac{{\partial \psi}}{{\partial r}}\right)
=\psi^3-\left(1-\xi^2\left(A_{\rm sc}+\frac{J}{r}\right)^2\right)\psi.
\end{equation}
The applied flux is now expressed explicitly in the equations by
\begin{equation}
J=\Phi_{ic}/\Phi_0,
\label{eq:DefofJ}
\end{equation}
and
$A_{\rm sc}(0)=A_{\rm sc}(\infty)=0$. For $r$ inside the SC, $\psi({r}) \ge 0$,  outside $\psi(r)=0$. The other boundary conditions are $\psi^\prime(r_{\rm in})=\psi^\prime(r_{\rm out})=0$. 
The analysis of Eqs.~\ref{eq:Asc} and \ref{eq:GL2} for the case $\xi  \ll \lambda \ll 1$ is described in Ref.~\cite{Gavish2020}.

\begin{figure}[tbph]
	\includegraphics[trim=0cm 1cm 0cm 0cm, clip=true,width=\columnwidth]{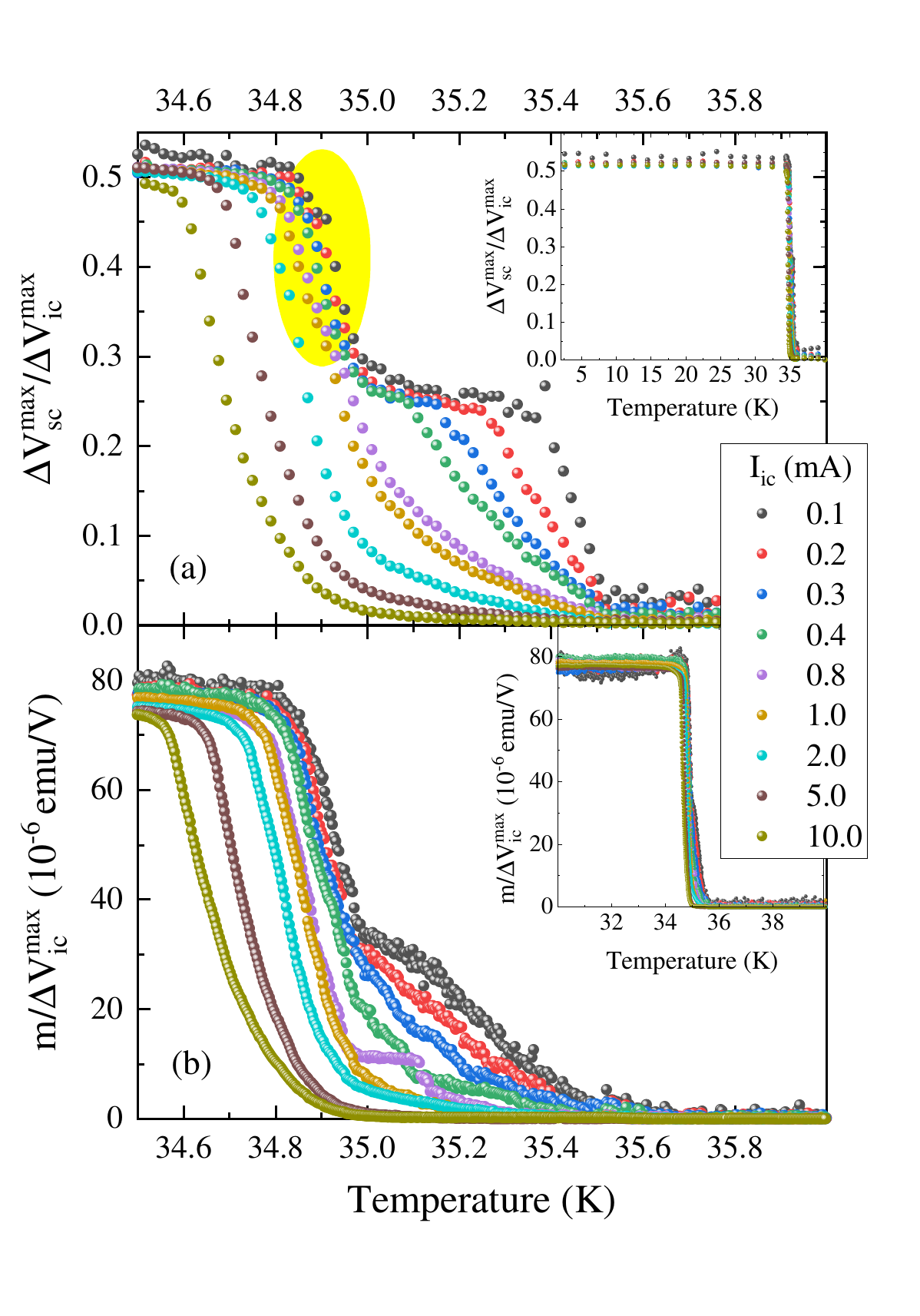}
	\caption{\textbf{Temperature dependence of normalized signals} (a) $\Delta V_{sc}^{max}/\Delta V_{ic}^{max}$ obtained by DC measurements (see Eq.~\ref{eq:Vratio_to_Aratio}) as a function of temperature close to the phase transition. The inset is a zoom-out on the entire temperature range. (b) The magnetic moment normalized by the coil signal (see Eq.~\ref{eq:m_to_A}) obtained by VSM measurements. Again, the inset is a zoom-out on a broader temperature range. }
		\label{fig:V_ration_vs_T}
	\end{figure}

The emerging picture is that when $J$ is small, the analysis of Sec.~\ref{Strong} is valid. Only for $J > {r_{in}^2}/{\sqrt 8 \xi \lambda }$, the order parameter's magnitude $\psi$ begins to diminish in the inner rim of the cylinder and the cylinder's hole is effectively larger than $r_{in}$. Nevertheless, the SC still expels the flux of the inner-coil and no critical point appears in $A_{sc}(R_{pl})$. The effective hole size $r_{in}^{eff}$ increases with increasing $J$, until $\psi$ survives only on a boundary layer of width $\lambda$ at $r_{out}$. At even larger $J$, the SC is no longer able to expel the applied flux, $A_{sc}$ does no longer grow with $I$, and vortices are expected to penetrate into the SC hole. These vortices are manifested in an increase of $\nabla \phi$. This behavior occurs at a folding point given by
\begin{equation}
J_{fold} \lesssim \frac{{r_{out}^2}}{{\sqrt 8 \xi \lambda }}.
\label{eq:jfold}
\end{equation}
The name ``folding'' means that increasing $J$ past $J_{fold}$ does not change the solution. The smaller $\xi$ and $\lambda$, the better the approximation of $J_{fold}$ is.

To evaluate the critical current $j_c$, we realize that when $j$ is pushed to a boundary layer of width $\lambda$ at $r_{out}$, it is still capable of expelling the inner-coil flux, but higher current will destroy SC completely. Therefore, ${\Phi _{ic}} = {\mu _0}j_c \lambda \pi r_{out}^2$. Using Eqs.~\ref{eq:DefofJ} and \ref{eq:jfold} we find 
\begin{equation}
{j_c} \gtrsim \frac{{{\Phi _0}}}{{\sqrt 8 \pi {\mu _0}{\lambda ^2}\xi }}
\label{eq:ciritical_current_density}
\end{equation}
where now $\lambda$ and $\xi$ are in units of length.

Although Eq.~\ref{eq:jfold} is derived for a tall cylinder we anticipate that it is valid for our ring. As long as $\lambda$ is smaller than all dimensions of the ring, currents will flow on the boundaries of the ``effective ring'', as in Fig.~\ref{Atot} and will be strongest at the inner rim of the ``effective ring'', but with a $J$ dependent $r_{in}^{eff}$. A change in behavior of the signal will take place only when $r_{in}^{eff} \simeq r_{out}-\lambda$ as in the cylinder case.

\section{Results} \label{Results}

Figure~\ref{fig:V_ration_vs_T}(a) depicts $\Delta V_{sc}^{max}/\Delta V_{ic}^{max}$ obtained by DC measurements. The signal is flat at low $T$ and drops close to $T_c$. As the current decreases, the drop of the signal is postponed to higher temperatures. At currents below $I=0.4$~mA a knee develops in the middle of the phase transition. Nevertheless, there is one $T_c=35.53$~K for all currents. Isolated islands of SC with stronger stiffness can not be the origin of these knees since only macroscopic closed lopes of SC can contribute to the signal. We speculate that these knees are related to SC surface states \cite{SamoilenkaPRB20}, with very small critical currents. In fact, knees were seen before in magnetization measurement on needle shaped LSCO, at very low fields, but they where not given much attention \cite{drachuck2012parallel}. The inset of Fig.~\ref{fig:V_ration_vs_T}(a) shows the full temperature range demonstrating that the normalized signal is independent of the applied coil current. In Fig.~\ref{fig:V_ration_vs_T}(b) we show the $\mathfrak{m}/\Delta V_{ic}^{max}$ data collected using the VSM method. Quantitatively, it looks the same as the DC measurement but less sharp and with few glitches of the signal. The knees disappear or smear and the uprise of the signal when cooling from $T_c$ is less abrupt. The inset again demonstrates that at low temperature the magnetic moment is proportional to the applied current as is mirrored in $\Delta V_{ic}^{max}$.

\begin{figure}[tbph]
	\includegraphics[trim=0cm 0cm 0cm 0cm, clip=true,width=\columnwidth]{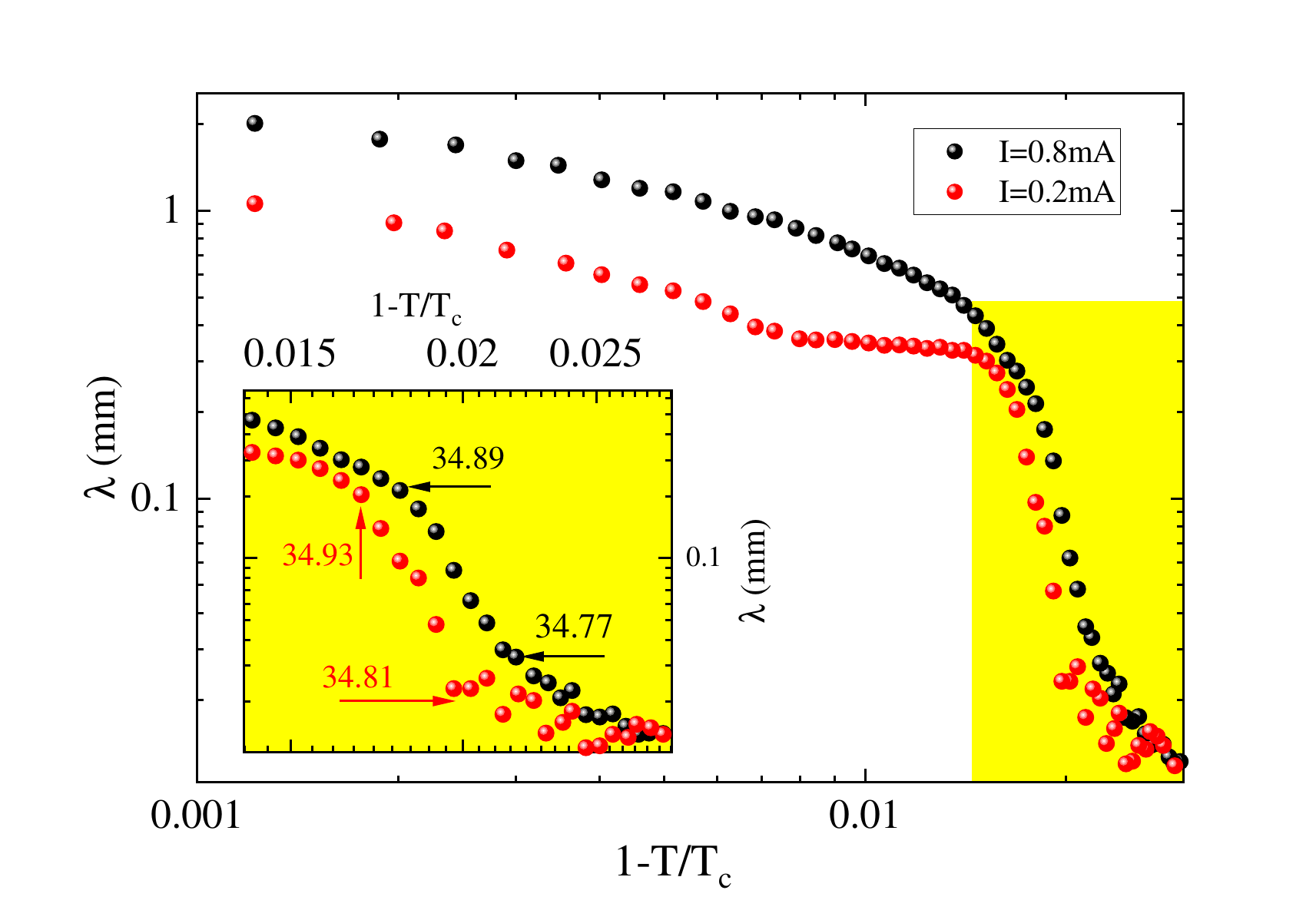}
	\caption{\textbf{Temperature dependence of the penetration depth.}  $\lambda$ extracted from the data of Fig.~\ref{fig:V_ration_vs_T}(a), based on Eq.~\ref{KGK}, over the full temperature range. The inset arrows mark $\lambda$'s that meet all criteria required for full Ginzburg-Landau analysis.}
	\label{fig:Lambda_vs_T}
\end{figure}

Using the measurements presented in Fig.~\ref{fig:V_ration_vs_T}(a), the experimentally determined conversion factor $G$, and the solution of Eq.~\ref{KGK} presented in Fig.~\ref{PDE}, we extract the penetration depth as if the solution is valid for all temperatures. The extracted $\lambda$ versus temperature with two applied currents $I=0.8$~mA and $I=0.2$~mA is depicted in Fig.~\ref{fig:Lambda_vs_T} on a log scale. Ideally we would like to find the $I \to 0$ limit of $\lambda$. However, at low temperatures where the signal saturates, the determination of $\lambda$ is noise. Close to $T_c$ there is a major behavior change at low current due to the knee. Moreover, a full Ginzburg-Landau analysis requires $\lambda \ll R_{pl}$. This leaves a small window where we can properly analyze our data. This window is marked by a yellow circle in Fig.~\ref{fig:V_ration_vs_T}(a), and by a yellow shade in Fig.~\ref{fig:Lambda_vs_T}. We zoom in on the shaded area in the inset of Fig.~\ref{fig:Lambda_vs_T} and show with arrows the temperature range where our analysis is valid.

\begin{figure}[tbph]
	\includegraphics[trim=0cm 0cm 0cm 0cm, clip=true,width=\columnwidth]{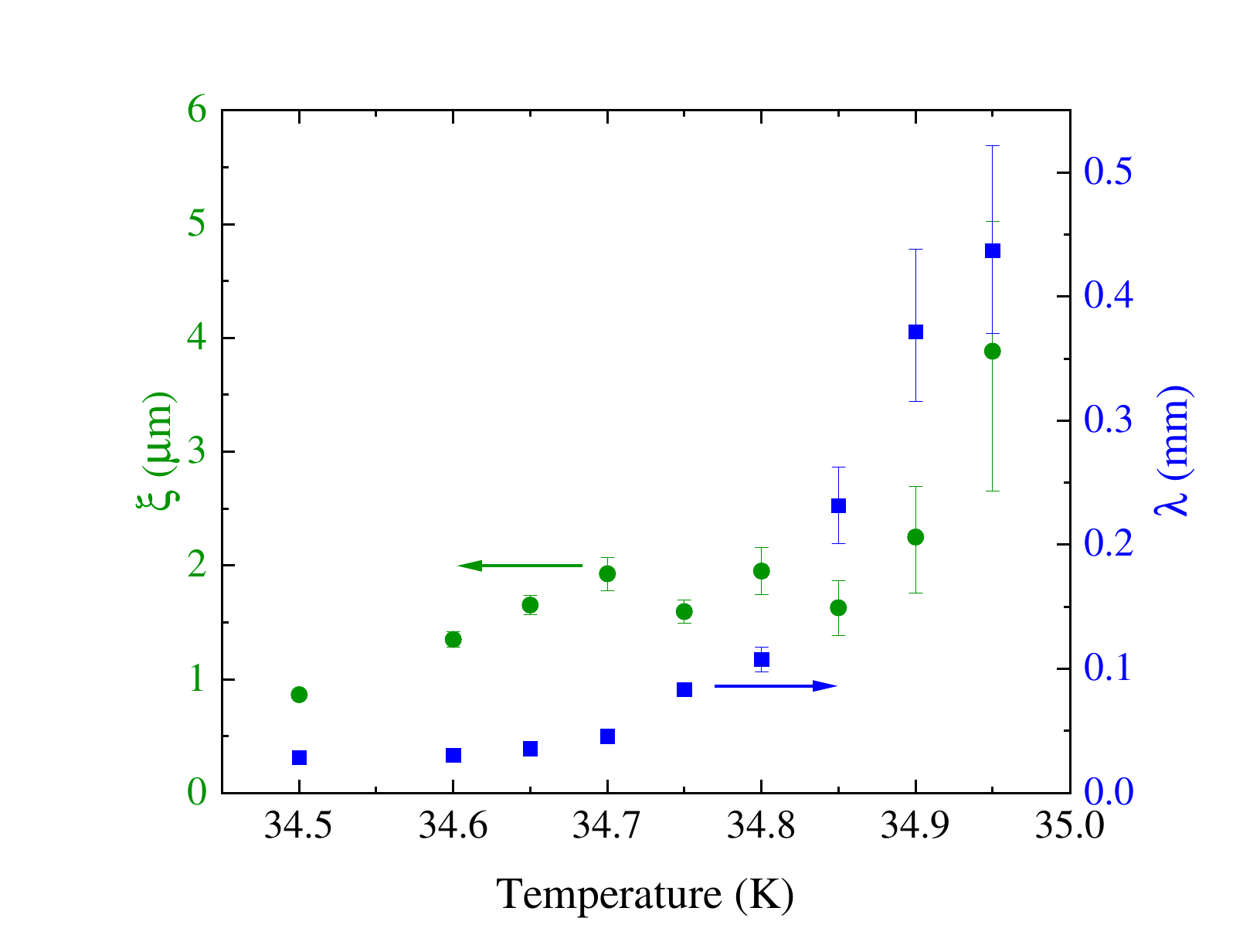}
	\caption{\textbf{Temperature dependence of the penetration depth and coherence length.} $\lambda$ (T) and $\xi$ (T) extracted from the data using the full Ginzburg-Landau analysis at a small temperature region where all approximations are valid and the Stiffnessometer is not saturated. }
	\label{fig:lambda_xi}
\end{figure}

As for $\xi$ and $j_c$; in Fig.~\ref{fig:CriticalCurrent} $\Delta V_{sc}^{max}(I)$ is measured at temperatures approaching $T_c$ but before the knee. We identify $I_c$ in this figure with $J_{fold}$ of Eq.~\ref{eq:jfold}. Calculating $\lambda$ at currents much lower than $I_c$, the flux generated by the coil at $I_c$ based on Fig.~\ref{BandAvsr}, and $J_{fold}$ from Eq.~\ref{eq:jfold} we extract $\xi$. The results for both $\lambda$ and $\xi$ are depicted in Fig.~\ref{fig:lambda_xi}. Since $\xi \ll \lambda$ there is a small temperature region where the Ginzburg-Landau analysis is self consistent. Using Eq.~\ref{eq:ciritical_current_density}, we find that the critical current density is on the order of $10^3$~Amm$^{-2}$ at the relevant temperature range, in agreement with measurements done in a field of $0.03$~T on similar samples \cite{wenEPL03}.

\section{Conclusions}

We demonstrated that the Stiffnessometer can measure penetration depth on a scale of millimeters, 
two orders of magnitude longer than ever before. This allows us to perform measurement closer to $T_c$ and explore the nature of the superconducting phase transition, or determine the stiffness at low $T$ in cases where it is naturally very weak as in thin films \cite{Nitsan}. The Stiffnessometer also allows measurements of very long coherence length $\xi$ on the order of micro-meters, equivalent to small critical current density on the order of  $10^3$~Amm$^{-2}$, properties which again are useful close to $T_c$. The measurements are done in a single apparatus, at zero magnetic field and with no leads, thus avoiding demagnetization, vortices, and out-of-equilibrium issues.

\section{Acknowledgments}

We are grateful for theoretical discussions with Assa Auerbach and Daniel Podolsky, mathematical assistance from Koby Rubinstein, and experimental tips from Ori Scaly. This study was financially supported by Israeli Science Foundation (ISF).


\begin{thebibliography}{19}%
	\makeatletter
	\providecommand \@ifxundefined [1]{%
		\@ifx{#1\undefined}
	}%
	\providecommand \@ifnum [1]{%
		\ifnum #1\expandafter \@firstoftwo
		\else \expandafter \@secondoftwo
		\fi
	}%
	\providecommand \@ifx [1]{%
		\ifx #1\expandafter \@firstoftwo
		\else \expandafter \@secondoftwo
		\fi
	}%
	\providecommand \natexlab [1]{#1}%
	\providecommand \enquote  [1]{``#1''}%
	\providecommand \bibnamefont  [1]{#1}%
	\providecommand \bibfnamefont [1]{#1}%
	\providecommand \citenamefont [1]{#1}%
	\providecommand \href@noop [0]{\@secondoftwo}%
	\providecommand \href [0]{\begingroup \@sanitize@url \@href}%
	\providecommand \@href[1]{\@@startlink{#1}\@@href}%
	\providecommand \@@href[1]{\endgroup#1\@@endlink}%
	\providecommand \@sanitize@url [0]{\catcode `\\12\catcode `\$12\catcode
		`\&12\catcode `\#12\catcode `\^12\catcode `\_12\catcode `\%12\relax}%
	\providecommand \@@startlink[1]{}%
	\providecommand \@@endlink[0]{}%
	\providecommand \url  [0]{\begingroup\@sanitize@url \@url }%
	\providecommand \@url [1]{\endgroup\@href {#1}{\urlprefix }}%
	\providecommand \urlprefix  [0]{URL }%
	\providecommand \Eprint [0]{\href }%
	\providecommand \doibase [0]{http://dx.doi.org/}%
	\providecommand \selectlanguage [0]{\@gobble}%
	\providecommand \bibinfo  [0]{\@secondoftwo}%
	\providecommand \bibfield  [0]{\@secondoftwo}%
	\providecommand \translation [1]{[#1]}%
	\providecommand \BibitemOpen [0]{}%
	\providecommand \bibitemStop [0]{}%
	\providecommand \bibitemNoStop [0]{.\EOS\space}%
	\providecommand \EOS [0]{\spacefactor3000\relax}%
	\providecommand \BibitemShut  [1]{\csname bibitem#1\endcsname}%
	\let\auto@bib@innerbib\@empty
	\bibitem [{\citenamefont {Tinkham}(2004)}]{tinkham2004introduction}%
	\BibitemOpen
	\bibfield  {author} {\bibinfo {author} {\bibfnamefont {M.}~\bibnamefont
			{Tinkham}},\ }\href@noop {} {\emph {\bibinfo {title} {Introduction to
				superconductivity}}}\ (\bibinfo  {publisher} {Courier Corporation},\ \bibinfo
	{year} {2004})\BibitemShut {NoStop}%
	\bibitem [{\citenamefont {Schrieffer}(2018)}]{schrieffer2018theory}%
	\BibitemOpen
	\bibfield  {author} {\bibinfo {author} {\bibfnamefont {J.~R.}\ \bibnamefont
			{Schrieffer}},\ }\href@noop {} {\emph {\bibinfo {title} {Theory of
				superconductivity}}}\ (\bibinfo  {publisher} {CRC Press},\ \bibinfo {year}
	{2018})\BibitemShut {NoStop}%
	\bibitem [{\citenamefont {De~Gennes}(2018)}]{de2018superconductivity}%
	\BibitemOpen
	\bibfield  {author} {\bibinfo {author} {\bibfnamefont {P.-G.}\ \bibnamefont
			{De~Gennes}},\ }\href@noop {} {\emph {\bibinfo {title} {Superconductivity of
				metals and alloys}}}\ (\bibinfo  {publisher} {CRC Press},\ \bibinfo {year}
	{2018})\BibitemShut {NoStop}%
	\bibitem [{\citenamefont {Uemura}\ \emph {et~al.}(1989)\citenamefont {Uemura},
		\citenamefont {Luke}, \citenamefont {Sternlieb}, \citenamefont {Brewer},
		\citenamefont {Carolan}, \citenamefont {Hardy}, \citenamefont {Kadono},
		\citenamefont {Kempton}, \citenamefont {Kiefl}, \citenamefont {Kreitzman}
		\emph {et~al.}}]{uemura1989universal}%
	\BibitemOpen
	\bibfield  {author} {\bibinfo {author} {\bibfnamefont {Y.}~\bibnamefont
			{Uemura}}, \bibinfo {author} {\bibfnamefont {G.}~\bibnamefont {Luke}},
		\bibinfo {author} {\bibfnamefont {B.}~\bibnamefont {Sternlieb}}, \bibinfo
		{author} {\bibfnamefont {J.}~\bibnamefont {Brewer}}, \bibinfo {author}
		{\bibfnamefont {J.}~\bibnamefont {Carolan}}, \bibinfo {author} {\bibfnamefont
			{W.}~\bibnamefont {Hardy}}, \bibinfo {author} {\bibfnamefont
			{R.}~\bibnamefont {Kadono}}, \bibinfo {author} {\bibfnamefont
			{J.}~\bibnamefont {Kempton}}, \bibinfo {author} {\bibfnamefont
			{R.}~\bibnamefont {Kiefl}}, \bibinfo {author} {\bibfnamefont
			{S.}~\bibnamefont {Kreitzman}},  \emph {et~al.},\ }\href@noop {} {\bibfield
		{journal} {\bibinfo  {journal} {Physical review letters}\ }\textbf {\bibinfo
			{volume} {62}},\ \bibinfo {pages} {2317} (\bibinfo {year}
		{1989})}\BibitemShut {NoStop}%
	\bibitem [{\citenamefont {Lamhot}\ \emph {et~al.}(2015)\citenamefont {Lamhot},
		\citenamefont {Yagil}, \citenamefont {Shapira}, \citenamefont {Kasahara},
		\citenamefont {Watashige}, \citenamefont {Shibauchi}, \citenamefont
		{Matsuda},\ and\ \citenamefont {Auslaender}}]{lamhot2015local}%
	\BibitemOpen
	\bibfield  {author} {\bibinfo {author} {\bibfnamefont {Y.}~\bibnamefont
			{Lamhot}}, \bibinfo {author} {\bibfnamefont {A.}~\bibnamefont {Yagil}},
		\bibinfo {author} {\bibfnamefont {N.}~\bibnamefont {Shapira}}, \bibinfo
		{author} {\bibfnamefont {S.}~\bibnamefont {Kasahara}}, \bibinfo {author}
		{\bibfnamefont {T.}~\bibnamefont {Watashige}}, \bibinfo {author}
		{\bibfnamefont {T.}~\bibnamefont {Shibauchi}}, \bibinfo {author}
		{\bibfnamefont {Y.}~\bibnamefont {Matsuda}}, \ and\ \bibinfo {author}
		{\bibfnamefont {O.~M.}\ \bibnamefont {Auslaender}},\ }\href@noop {}
	{\bibfield  {journal} {\bibinfo  {journal} {Physical Review B}\ }\textbf
		{\bibinfo {volume} {91}},\ \bibinfo {pages} {060504} (\bibinfo {year}
		{2015})}\BibitemShut {NoStop}%
	\bibitem [{\citenamefont {Morenzoni}\ \emph {et~al.}(2002)\citenamefont
		{Morenzoni}, \citenamefont {Gl{\"u}ckler}, \citenamefont {Prokscha},
		\citenamefont {Khasanov}, \citenamefont {Luetkens}, \citenamefont {Birke},
		\citenamefont {Forgan}, \citenamefont {Niedermayer},\ and\ \citenamefont
		{Pleines}}]{morenzoni2002implantation}%
	\BibitemOpen
	\bibfield  {author} {\bibinfo {author} {\bibfnamefont {E.}~\bibnamefont
			{Morenzoni}}, \bibinfo {author} {\bibfnamefont {H.}~\bibnamefont
			{Gl{\"u}ckler}}, \bibinfo {author} {\bibfnamefont {T.}~\bibnamefont
			{Prokscha}}, \bibinfo {author} {\bibfnamefont {R.}~\bibnamefont {Khasanov}},
		\bibinfo {author} {\bibfnamefont {H.}~\bibnamefont {Luetkens}}, \bibinfo
		{author} {\bibfnamefont {M.}~\bibnamefont {Birke}}, \bibinfo {author}
		{\bibfnamefont {E.}~\bibnamefont {Forgan}}, \bibinfo {author} {\bibfnamefont
			{C.}~\bibnamefont {Niedermayer}}, \ and\ \bibinfo {author} {\bibfnamefont
			{M.}~\bibnamefont {Pleines}},\ }\href@noop {} {\bibfield  {journal} {\bibinfo
			{journal} {Nuclear Instruments and Methods in Physics Research Section B:
				Beam Interactions with Materials and Atoms}\ }\textbf {\bibinfo {volume}
			{192}},\ \bibinfo {pages} {254} (\bibinfo {year} {2002})}\BibitemShut
	{NoStop}%
	\bibitem [{\citenamefont {Morenzoni}\ \emph {et~al.}(2004)\citenamefont
		{Morenzoni}, \citenamefont {Prokscha}, \citenamefont {Suter}, \citenamefont
		{Luetkens},\ and\ \citenamefont {Khasanov}}]{morenzoni2004nano}%
	\BibitemOpen
	\bibfield  {author} {\bibinfo {author} {\bibfnamefont {E.}~\bibnamefont
			{Morenzoni}}, \bibinfo {author} {\bibfnamefont {T.}~\bibnamefont {Prokscha}},
		\bibinfo {author} {\bibfnamefont {A.}~\bibnamefont {Suter}}, \bibinfo
		{author} {\bibfnamefont {H.}~\bibnamefont {Luetkens}}, \ and\ \bibinfo
		{author} {\bibfnamefont {R.}~\bibnamefont {Khasanov}},\ }\href@noop {}
	{\bibfield  {journal} {\bibinfo  {journal} {Journal of Physics: Condensed
				Matter}\ }\textbf {\bibinfo {volume} {16}},\ \bibinfo {pages} {S4583}
		(\bibinfo {year} {2004})}\BibitemShut {NoStop}%
	\bibitem [{\citenamefont {Prozorov}\ and\ \citenamefont
		{Giannetta}(2006)}]{prozorov2006magnetic}%
	\BibitemOpen
	\bibfield  {author} {\bibinfo {author} {\bibfnamefont {R.}~\bibnamefont
			{Prozorov}}\ and\ \bibinfo {author} {\bibfnamefont {R.~W.}\ \bibnamefont
			{Giannetta}},\ }\href@noop {} {\bibfield  {journal} {\bibinfo  {journal}
			{Superconductor Science and Technology}\ }\textbf {\bibinfo {volume} {19}},\
		\bibinfo {pages} {R41} (\bibinfo {year} {2006})}\BibitemShut {NoStop}%
	\bibitem [{\citenamefont {Drachuck}\ \emph {et~al.}(2012)\citenamefont
		{Drachuck}, \citenamefont {Shay}, \citenamefont {Bazalitsky}, \citenamefont
		{Berger},\ and\ \citenamefont {Keren}}]{drachuck2012parallel}%
	\BibitemOpen
	\bibfield  {author} {\bibinfo {author} {\bibfnamefont {G.}~\bibnamefont
			{Drachuck}}, \bibinfo {author} {\bibfnamefont {M.}~\bibnamefont {Shay}},
		\bibinfo {author} {\bibfnamefont {G.}~\bibnamefont {Bazalitsky}}, \bibinfo
		{author} {\bibfnamefont {J.}~\bibnamefont {Berger}}, \ and\ \bibinfo {author}
		{\bibfnamefont {A.}~\bibnamefont {Keren}},\ }\href@noop {} {\bibfield
		{journal} {\bibinfo  {journal} {Physical Review B}\ }\textbf {\bibinfo
			{volume} {85}},\ \bibinfo {pages} {184518} (\bibinfo {year}
		{2012})}\BibitemShut {NoStop}%
	\bibitem [{\citenamefont {Zhou}\ \emph {et~al.}(2007)\citenamefont {Zhou},
		\citenamefont {Maiorov}, \citenamefont {Wang}, \citenamefont
		{MacManus-Driscoll}, \citenamefont {Holesinger}, \citenamefont {Civale},
		\citenamefont {Jia},\ and\ \citenamefont {Foltyn}}]{zhou2007improved}%
	\BibitemOpen
	\bibfield  {author} {\bibinfo {author} {\bibfnamefont {H.}~\bibnamefont
			{Zhou}}, \bibinfo {author} {\bibfnamefont {B.}~\bibnamefont {Maiorov}},
		\bibinfo {author} {\bibfnamefont {H.}~\bibnamefont {Wang}}, \bibinfo {author}
		{\bibfnamefont {J.}~\bibnamefont {MacManus-Driscoll}}, \bibinfo {author}
		{\bibfnamefont {T.}~\bibnamefont {Holesinger}}, \bibinfo {author}
		{\bibfnamefont {L.}~\bibnamefont {Civale}}, \bibinfo {author} {\bibfnamefont
			{Q.}~\bibnamefont {Jia}}, \ and\ \bibinfo {author} {\bibfnamefont
			{S.}~\bibnamefont {Foltyn}},\ }\href@noop {} {\bibfield  {journal} {\bibinfo
			{journal} {Superconductor Science and Technology}\ }\textbf {\bibinfo
			{volume} {21}},\ \bibinfo {pages} {025001} (\bibinfo {year}
		{2007})}\BibitemShut {NoStop}%
	\bibitem [{\citenamefont {Shay}\ \emph {et~al.}(2009)\citenamefont {Shay},
		\citenamefont {Keren}, \citenamefont {Koren}, \citenamefont {Kanigel},
		\citenamefont {Shafir}, \citenamefont {Marcipar}, \citenamefont
		{Nieuwenhuys}, \citenamefont {Morenzoni}, \citenamefont {Suter},
		\citenamefont {Prokscha} \emph {et~al.}}]{shay2009interaction}%
	\BibitemOpen
	\bibfield  {author} {\bibinfo {author} {\bibfnamefont {M.}~\bibnamefont
			{Shay}}, \bibinfo {author} {\bibfnamefont {A.}~\bibnamefont {Keren}},
		\bibinfo {author} {\bibfnamefont {G.}~\bibnamefont {Koren}}, \bibinfo
		{author} {\bibfnamefont {A.}~\bibnamefont {Kanigel}}, \bibinfo {author}
		{\bibfnamefont {O.}~\bibnamefont {Shafir}}, \bibinfo {author} {\bibfnamefont
			{L.}~\bibnamefont {Marcipar}}, \bibinfo {author} {\bibfnamefont
			{G.}~\bibnamefont {Nieuwenhuys}}, \bibinfo {author} {\bibfnamefont
			{E.}~\bibnamefont {Morenzoni}}, \bibinfo {author} {\bibfnamefont
			{A.}~\bibnamefont {Suter}}, \bibinfo {author} {\bibfnamefont
			{T.}~\bibnamefont {Prokscha}},  \emph {et~al.},\ }\href@noop {} {\bibfield
		{journal} {\bibinfo  {journal} {Physical Review B}\ }\textbf {\bibinfo
			{volume} {80}},\ \bibinfo {pages} {144511} (\bibinfo {year}
		{2009})}\BibitemShut {NoStop}%
	\bibitem [{\citenamefont {Talantsev}\ \emph {et~al.}(2014)\citenamefont
		{Talantsev}, \citenamefont {Strickland}, \citenamefont {Wimbush},
		\citenamefont {Storey}, \citenamefont {Tallon},\ and\ \citenamefont
		{Long}}]{talantsev2014hole}%
	\BibitemOpen
	\bibfield  {author} {\bibinfo {author} {\bibfnamefont {E.}~\bibnamefont
			{Talantsev}}, \bibinfo {author} {\bibfnamefont {N.}~\bibnamefont
			{Strickland}}, \bibinfo {author} {\bibfnamefont {S.}~\bibnamefont {Wimbush}},
		\bibinfo {author} {\bibfnamefont {J.}~\bibnamefont {Storey}}, \bibinfo
		{author} {\bibfnamefont {J.}~\bibnamefont {Tallon}}, \ and\ \bibinfo {author}
		{\bibfnamefont {N.}~\bibnamefont {Long}},\ }\href@noop {} {\bibfield
		{journal} {\bibinfo  {journal} {Applied Physics Letters}\ }\textbf {\bibinfo
			{volume} {104}},\ \bibinfo {pages} {242601} (\bibinfo {year}
		{2014})}\BibitemShut {NoStop}%
	\bibitem [{\citenamefont {Talantsev}\ and\ \citenamefont
		{Tallon}(2015)}]{talantsev2015universal}%
	\BibitemOpen
	\bibfield  {author} {\bibinfo {author} {\bibfnamefont {E.~F.}\ \bibnamefont
			{Talantsev}}\ and\ \bibinfo {author} {\bibfnamefont {J.~L.}\ \bibnamefont
			{Tallon}},\ }\href@noop {} {\bibfield  {journal} {\bibinfo  {journal} {Nature
				communications}\ }\textbf {\bibinfo {volume} {6}},\ \bibinfo {pages} {1}
		(\bibinfo {year} {2015})}\BibitemShut {NoStop}%
	\bibitem [{\citenamefont {Gavish}\ \emph {et~al.}()\citenamefont {Gavish},
		\citenamefont {Kenneth},\ and\ \citenamefont {Keren}}]{Gavish2020}%
	\BibitemOpen
	\bibfield  {author} {\bibinfo {author} {\bibfnamefont {N.}~\bibnamefont
			{Gavish}}, \bibinfo {author} {\bibfnamefont {O.}~\bibnamefont {Kenneth}}, \
		and\ \bibinfo {author} {\bibfnamefont {A.}~\bibnamefont {Keren}},\
	}\href@noop {} {\bibinfo  {journal} {arXiv:2005.11686}\ }\BibitemShut
	{NoStop}%
	\bibitem [{\citenamefont {Kapon}\ \emph {et~al.}(2019)\citenamefont {Kapon},
		\citenamefont {Salman}, \citenamefont {Mangel}, \citenamefont {Prokscha},
		\citenamefont {Gavish},\ and\ \citenamefont {Keren}}]{kapon2019phase}%
	\BibitemOpen
	\bibfield  {journal} {  }\bibfield  {author} {\bibinfo {author} {\bibfnamefont
			{I.}~\bibnamefont {Kapon}}, \bibinfo {author} {\bibfnamefont
			{Z.}~\bibnamefont {Salman}}, \bibinfo {author} {\bibfnamefont
			{I.}~\bibnamefont {Mangel}}, \bibinfo {author} {\bibfnamefont
			{T.}~\bibnamefont {Prokscha}}, \bibinfo {author} {\bibfnamefont
			{N.}~\bibnamefont {Gavish}}, \ and\ \bibinfo {author} {\bibfnamefont
			{A.}~\bibnamefont {Keren}},\ }\href@noop {} {\bibfield  {journal} {\bibinfo
			{journal} {Nature communications}\ }\textbf {\bibinfo {volume} {10}},\
		\bibinfo {pages} {1} (\bibinfo {year} {2019})}\BibitemShut {NoStop}%
	\bibitem [{\citenamefont {Hecht}(2012)}]{freefem}%
	\BibitemOpen
	\bibfield  {author} {\bibinfo {author} {\bibfnamefont {F.}~\bibnamefont
			{Hecht}},\ }\href@noop {} {\bibfield  {journal} {\bibinfo  {journal} {J.
				Numer. Math.}\ }\textbf {\bibinfo {volume} {20}},\ \bibinfo {pages} {251}
		(\bibinfo {year} {2012})}\BibitemShut {NoStop}%
	\bibitem [{\citenamefont {Samoilenka}\ and\ \citenamefont
		{Babaev}(2020)}]{SamoilenkaPRB20}%
	\BibitemOpen
	\bibfield  {author} {\bibinfo {author} {\bibfnamefont {A.}~\bibnamefont
			{Samoilenka}}\ and\ \bibinfo {author} {\bibfnamefont {E.}~\bibnamefont
			{Babaev}},\ }\href {\doibase 10.1103/PhysRevB.101.134512} {\bibfield
		{journal} {\bibinfo  {journal} {Phys. Rev. B}\ }\textbf {\bibinfo {volume}
			{101}},\ \bibinfo {pages} {134512} (\bibinfo {year} {2020})}\BibitemShut
	{NoStop}%
	\bibitem [{\citenamefont {Wen}\ \emph {et~al.}(2003)\citenamefont {Wen},
		\citenamefont {Yang}, \citenamefont {Li}, \citenamefont {Zeng}, \citenamefont
		{Soukiassian}, \citenamefont {Si},\ and\ \citenamefont {Xi}}]{wenEPL03}%
	\BibitemOpen
	\bibfield  {author} {\bibinfo {author} {\bibfnamefont {H.}~\bibnamefont
			{Wen}}, \bibinfo {author} {\bibfnamefont {H.}~\bibnamefont {Yang}}, \bibinfo
		{author} {\bibfnamefont {S.}~\bibnamefont {Li}}, \bibinfo {author}
		{\bibfnamefont {X.}~\bibnamefont {Zeng}}, \bibinfo {author} {\bibfnamefont
			{A.}~\bibnamefont {Soukiassian}}, \bibinfo {author} {\bibfnamefont
			{W.}~\bibnamefont {Si}}, \ and\ \bibinfo {author} {\bibfnamefont
			{X.}~\bibnamefont {Xi}},\ }\href@noop {} {\bibfield  {journal} {\bibinfo
			{journal} {EPL (Europhysics Letters)}\ }\textbf {\bibinfo {volume} {64}},\
		\bibinfo {pages} {790} (\bibinfo {year} {2003})}\BibitemShut {NoStop}%
	\bibitem [{\citenamefont {Blau}()}]{Nitsan}%
	\BibitemOpen
	\bibfield  {author} {\bibinfo {author} {\bibfnamefont {N.}~\bibnamefont
			{Blau}},\ }\href@noop {} {\bibinfo  {journal} {In preparation}\ }\BibitemShut
	{NoStop}%
\end{thebibliography}
%

\end{document}